# Critical Role of Quantum Dynamical Effects in the Raman Spectroscopy of Liquid Water

Xinzijian Liu[1] and Jian Liu[1, †]

1. Beijing National Laboratory for Molecular Sciences, Institute of Theoretical and Computational Chemistry, College of Chemistry and Molecular Engineering, Peking University, Beijing 100871, China

† Electronic mail: jianliupku@pku.edu.cn

**Abstract**

Understanding the Raman spectroscopy at the atomistic level is important for the elucidation of dynamical processes in liquid water. Because the polarizability (or its time derivative) is often a highly nonlinear function of coordinates or/and momenta, we employ the linearized semiclassical initial value representation for quantum dynamical simulations of liquid water (and heavy water) under ambient conditions based on an *ab initio* based, flexible, polarizable model (the POLI2VS force field). It is shown that quantum dynamical effects play a critical role in reproducing the peaks in the intermediate region between the librational and bending bands, those between the bending and stretching bands, and the double-peak in the stretching band in the experimental isotropic Raman spectrum. In contrast, quantum dynamical effects are important but less decisive in the anisotropic Raman spectrum. By selectively freezing either the intramolecular O-H stretching or H-O-H bending mode, we demonstrate that the peak in the intermediate region (2000-2400 $cm^{-1}$) of the isotropic Raman spectrum arises from the interplay of the stretching and bending motions while a substantial part of the peak in the same intermediate region of the anisotropic Raman spectrum may be attributed to the combined motion of the bending and librational modes.

I.   Introduction

Liquid water plays a crucial role in various phenomena/processes in chemistry, physics, biology, geology, climate research, etc.[1-13]. This is mainly related to its flexible and dynamic hydrogen-bonding networks[14-18]. The significance of nuclear quantum effects on the properties of liquid water has not yet been fully resolved, although important progress has been made in last two decades[19-43] by virtue of recent development on path integral molecular dynamics (PIMD) techniques [such as reversible multiple time scale method[44], ring polymer contraction[45], color noise methods[46, 47], and more generally applicable "middle" thermostat scheme[48-51]] for structural and thermodynamic properties and on approximate quantum dynamics methods [including linearized semiclassical initial value representation (LSC-IVR)[52-59], derivative forward-backward semiclassical dynamics (FBSD)[60-64], centroid molecular dynamics (CMD)[65-68], ring polymer molecular dynamics (RPMD)[69-72], etc.] for dynamical properties.

Since water vibrations have a complex mixed character of both intramolecular dynamics and the fluctuating hydrogen bond network, vibrational spectroscopy often serves as a probe of structure and dynamics in liquid water[2, 73]. While the infra-red (IR) response is linear, the Raman response is of the third order. In pure liquid water the Raman spectrum demonstrates more pronounced features than the IR spectrum[2]. For instance, while a more distinct peak between the bending and stretching bands appears around 2061 cm$^{-1}$ in the isotropic Raman spectrum and around 2110 cm$^{-1}$ in the anisotropic Raman spectrum, two peaks rather than one exist in the O-H stretching band of the isotropic Raman spectrum. It has long been suspected whether the Fermi resonance plays a role in the Raman spectrum of ambient water[2, 74-76].

In the present paper we focus on quantum dynamical effects in anisotropic and isotropic Raman spectra of liquid water. The isotropic (or polarized) component of the differential scattering cross section for scattering into a frequency range $d\omega$ and a solid angle $d\Omega$ is related to the Fourier transform of the standard version of the isotropic polarizability auto-correlation function by

$$\frac{d^2\sigma_{iso}}{d\omega d\Omega} \propto (\omega_0 - \omega)^4 \frac{1}{2\pi} \int_{-\infty}^{+\infty} e^{-i\omega t} \langle \hat{\alpha}_{iso}(0)\hat{\alpha}_{iso}(t) \rangle_{std} dt \quad , \tag{1}$$

with $\omega_0$ as the circular frequency of the excitation light, and the isotropic polarizability defined by $\alpha_{iso} \equiv \text{Tr}(\boldsymbol{\alpha})/3$, where $\boldsymbol{\alpha}$ represents the $3 \times 3$ collective polarizability tensor of the system[77]. Similarly, the anisotropic (or depolarized) component is connected to the Fourier transform of the standard version of the anisotropic polarizability auto-correlation function by

$$\frac{d^2\sigma_{aniso}}{d\omega d\Omega} \propto (\omega_0 - \omega)^4 \frac{1}{2\pi} \int_{-\infty}^{+\infty} e^{-i\omega t} \langle \text{Tr}[\hat{\boldsymbol{\beta}}(0)\hat{\boldsymbol{\beta}}(t)] \rangle_{std} dt \tag{2}$$

with the anisotropic polarizability defined by $\boldsymbol{\beta} \equiv \boldsymbol{\alpha} - \alpha_{iso}\mathbf{1}$.[77] The reduced Raman intensity $R(\omega)$ defined by Lund *et al.* [76, 78, 79] is given by

$$R(\omega) \propto (\omega_0 - \omega)^{-4} \omega \tanh\left(\frac{\beta\hbar\omega}{2}\right) \frac{d^2\sigma}{d\omega d\Omega} \quad . \tag{3}$$

The reduced isotropic Raman spectrum is then produced by

$$R_{iso}(\omega) \propto \frac{\omega(1-e^{-\beta\hbar\omega})}{2\pi(1+e^{-\beta\hbar\omega})} \int_{-\infty}^{+\infty} e^{-i\omega t} \langle \hat{\alpha}_{iso}(0)\hat{\alpha}_{iso}(t)\rangle_{std} \, dt \quad, \tag{4}$$

which may be recast in terms of the real part of the standard version of the isotropic polarizability-derivative auto-correlation function as

$$R_{iso}(\omega) \propto \frac{1-e^{-\beta\hbar\omega}}{\pi\omega(1+e^{-\beta\hbar\omega})^2} \int_{-\infty}^{+\infty} e^{-i\omega t} \, \mathrm{Re}\langle \hat{\dot{\alpha}}_{iso}(0)\hat{\dot{\alpha}}_{iso}(t)\rangle_{std} \, dt \quad. \tag{5}$$

Here $\dot{\alpha}_{iso}$ is the derivative of $\alpha_{iso}$ over time. Similarly, the reduced anisotropic Raman spectrum is related to the Fourier transform of the real part of the standard version of the anisotropic polarizability-derivative auto-correlation function,

$$R_{aniso}(\omega) \propto \frac{1-e^{-\beta\hbar\omega}}{\pi\omega(1+e^{-\beta\hbar\omega})^2} \int_{-\infty}^{+\infty} e^{-i\omega t} \, \mathrm{Re}\langle \mathrm{Tr}[\hat{\dot{\boldsymbol{\beta}}}(0)\hat{\dot{\boldsymbol{\beta}}}(t)]\rangle_{std} \, dt \quad, \tag{6}$$

where the anisotropic polarizability-derivative $\dot{\boldsymbol{\beta}}$ is the time derivative of $\boldsymbol{\beta}$. As suggested by Eq. (5) and Eq. (6), the isotropic and anisotropic Raman spectra imply the scalar and vector nature of the field-matter interactions, respectively. One can also express Eq. (5) and Eq. (6) in terms of the Kubo-transformed version of the correlation function, i.e.,

$$R_{iso}(\omega) \propto \frac{\beta\hbar}{2\pi} \frac{1}{1+e^{-\beta\hbar\omega}} \int_{-\infty}^{+\infty} e^{-i\omega t} \langle \hat{\dot{\alpha}}_{iso}(0)\hat{\dot{\alpha}}_{iso}(t)\rangle_{Kubo} \, dt \tag{7}$$

and

$$R_{\text{aniso}}(\omega) \propto \frac{\beta\hbar}{2\pi} \frac{1}{1+e^{-\beta\hbar\omega}} \int_{-\infty}^{+\infty} e^{-i\omega t} \left\langle \text{Tr}\left[\hat{\boldsymbol{\beta}}(0)\hat{\boldsymbol{\beta}}(t)\right] \right\rangle_{\text{Kubo}} dt . \qquad (8)$$

A more compact form of the time correlation functions[77, 80, 81] listed in Eqs. (1)-(2) and in Eqs. (4)-(8) may be given by

$$\langle \hat{A}(0)\hat{B}(t) \rangle \equiv C_{AB}(t) = \frac{1}{Z} \text{Tr}\left(\hat{A}^{\beta} e^{i\hat{H}t/\hbar} \hat{B} e^{-i\hat{H}t/\hbar}\right) \qquad (9)$$

where $\hat{A}_{\text{std}}^{\beta} = e^{-\beta\hat{H}}\hat{A}$ for the standard version of the correlation function, or $\hat{A}_{\text{sym}}^{\beta} = e^{-\beta\hat{H}/2}\hat{A}e^{-\beta\hat{H}/2}$ for the symmetrized version[82], or $\hat{A}_{\text{Kubo}}^{\beta} = \frac{1}{\beta}\int_0^{\beta} d\lambda \, e^{-(\beta-\lambda)\hat{H}} \hat{A} e^{-\lambda\hat{H}}$ for the Kubo-transformed version[83]. These three versions are related to one another by the following identities between their Fourier transforms,

$$\frac{\beta\hbar\omega}{1-e^{-\beta\hbar\omega}} I_{AB}^{\text{Kubo}}(\omega) = I_{AB}^{\text{std}}(\omega) = e^{\beta\hbar\omega/2} I_{AB}^{\text{sym}}(\omega) \qquad (10)$$

where $I_{AB}(\omega) = \int_{-\infty}^{\infty} dt \, e^{-i\omega t} C_{AB}(t)$ etc. Here $Z = \text{Tr}\left(e^{-\beta\hat{H}}\right)$ ($\beta = 1/k_B T$) is the partition function and $\hat{H}$ the (time-independent) Hamiltonian of the system, and $\hat{A}$ and $\hat{B}$ are relevant operators.

An important component for simulating the Raman spectroscopy of liquid water in the entire frequency domain (from the librational band to the O-H stretching band) is the real time quantum dynamics method. The mixed quantum mechanics/classical mechanics approach often focuses on the stretching band by decoupling the O-H stretch from the other degrees of freedom[26, 84]. Although this type of approach often offers instructive information for the stretching band, it completely ignores the interplay of different modes and is not able to describe the entire spectrum. The two approximate quantum dynamics methods—CMD and RPMD are two possible candidates and have recently been used to compute the Raman spectroscopy[40, 42]. Note that the polarizability (or polarizability-derivative) of water molecules is often a (highly) nonlinear operator (i.e., nonlinear function of nuclear degrees of freedom). It is not well justified that these versions of CMD[65-68] and RPMD[69-72] are good for calculating the polarizability (or polarizability-derivative) correlation function, because these approximate quantum dynamics methods are incapable of giving exact correlation functions of nonlinear operators even for the harmonic system[85-94]. While the association band around 2110 cm$^{-1}$ is missing in the CMD anisotropic spectrum in Ref. [40], only one peak (instead of two peaks) appears in the O-H stretching band of the CMD (or thermostatted RPMD) isotropic spectrum in Ref. [42]. A more reasonable candidate for calculating the Raman spectroscopy may be the linearized semiclassical initial value representation (LSC-IVR) of Miller *et al.*[28, 52-55, 57, 59] (or the classical Wigner model for the correlation function), because it treats both linear and *nonlinear* operators equally well and recover exact correlation functions in the classical $(\hbar \to 0)$, high temperature $(\beta \to 0)$, and harmonic limits.

In addition to the real time quantum dynamics method, the potential energy, dipole moment, and polarizability tensor surfaces (PES, DMS, and PTS) are another essential component for studying the Raman spectroscopy of liquid water. It is crucial to employ either *ab initio* calculations on the fly or *ab*

*initio*-based force fields to avoid the "double counting problem" for quantum simulations based on conventional classical water force fields[23, 24, 27, 28, 30, 33]. Significant progress has been made in developing force fields or electronic structure methods for the accurate description of liquid water[25, 95-114]. The water force field employed in the paper is a polarizable, flexible, and transferable potential for inter- and intramolecular vibrational spectroscopy (POLI2VS) developed by Hasegawa and Tanimura[115] based on the POLIR potential[116] with atomic multipoles. In the POLI2VS force field the distributed multipole analysis and the distributed polarization model proposed by Stone and Alderton[117, 118] were implemented to mimic the instantaneous charge density of a water molecule in a consistent way for describing the dipole moment surface and the polarization effect[115]. While multipoles up to quadrupole are used for the permanent part, multipoles up to dipole are employed for the induced part[115]. Comparing to the experimental data[76], the Raman spectrum simulated by classical molecular dynamics (MD) misses some important structures in the intermediate region between the librational and bending bands and in the intermediate region between the bending and stretching bands, in addition that classical MD produces a blue-shifted and more narrow stretching band[115]. Such deviations should be corrected/alleviated if quantum dynamical effects are considered. Investigation on this can shed light on the role of nuclear quantum dynamical effects in Raman modes of liquid water. We note that quantum dynamical effects have already been demonstrated to play an important role in the vibrational spectroscopy of liquid water by a comprehensive study on the IR spectrum[33] and a heuristic illustration with a model Hamiltonian on linear and nonlinear spectra[119].

The paper is organized as follows: Section II first briefly reviews the LSC-IVR methodology for time correlation functions. Section III then gives the explicit formulations for the isotropic and anisotropic polarizability-derivative correlation functions for the corresponding Raman spectra. Section IV studies the isotropic and anisotropic Raman spectra of liquid water and heavy water at 298.15K, followed by further analyses and discussions. A brief summary and conclusion remarks are presented in Section V.

## II.     Simulation methodology

### A. Linearized semiclassical initial value representation for the correlation function

Simulations of time correlation functions in Eqs. (4)-(9) for large molecular systems often present a challenge. While classical molecular dynamics (MD) fails to capture quantum effects, exact interpretations of quantum mechanics (such as the time-dependent Schrödinger equation and Feynman's real time path integral) are impractical to apply to liquid water. A practical approximate quantum dynamics method is the LSC-IVR or classical Wigner model for the correlation function in Eq. (9), i.e.,

$$C_{AB}^{\text{LSC-IVR}}(t) = Z^{-1}(2\pi\hbar)^{-3N} \int d\mathbf{x}_0 \int d\mathbf{p}_0 A_W^\beta(\mathbf{x}_0, \mathbf{p}_0) B_W(\mathbf{x}_t, \mathbf{p}_t) \qquad (11)$$

where $A_W^\beta$ and $B_W$ are the Wigner functions[120] corresponding to these operators,

$$O_W(\mathbf{x},\mathbf{p}) = \int d\Delta\mathbf{x} \langle \mathbf{x} - \Delta\mathbf{x}/2 | \hat{O} | \mathbf{x} + \Delta\mathbf{x}/2 \rangle e^{i\Delta\mathbf{x}^T \mathbf{p}/\hbar} \qquad (12)$$

for any operator $\hat{O}$. Here $N$ is the number of particles in the system, and $(\mathbf{x}_0, \mathbf{p}_0)$ is the set of initial conditions (i.e., coordinates and momenta) for a classical trajectory, $(\mathbf{x}_t(\mathbf{x}_0, \mathbf{p}_0), \mathbf{p}_t(\mathbf{x}_0, \mathbf{p}_0))$ being the phase point at time $t$ along this trajectory.

It was originally derived from the initial value representations (IVR) of semiclassical (SC) theory[121, 122], which approximates the forward (backward) time evolution operator $e^{-i\hat{H}t/\hbar}$ ($e^{i\hat{H}t/\hbar}$) by a phase space average over the initial conditions of forward (backward) classical trajectories[121, 123-125]. The LSC-IVR was obtained by making the approximation that the dominant contribution to the phase space averages comes from forward and backward trajectories that are infinitesimally close to one another, and then linearizing the difference between the forward and backward actions (and other quantities in the integrand)[52-54]. Other approximate routes[55-57, 126, 127] have also been proposed to derive the LSC-IVR/classical Wigner model for correlation functions (other than simply postulating it). It was shown that the exact quantum time correlation function can be expressed in the same form as Eq. (11), with an associated dynamics in the single phase space[126, 127], and it was furthermore demonstrated that the LSC-IVR is its classical limit ($\hbar \to 0$), high temperature limit ($\beta \to 0$), and harmonic limit (even for correlation functions involving nonlinear operators). It should be stressed that all these approximate routes also indicate that the LSC-IVR is the short time approximation ($t \to 0$) to the quantum time correlation function, which has been well demonstrated by the perturbation theory[128] and by the maximum entropy analytical continuation test and other analyses for condensed phase systems[58, 129-131].

The LSC-IVR has two drawbacks. One is that it cannot describe true quantum coherence effects in time correlation functions. Such effects are expected to be quenched and not important in condensed phase systems such as liquid water since too many degrees of freedom are involved[28, 33, 124, 125]. The other drawback is that the distribution generated for the operator $\hat{A}^\beta$ for the case $\hat{A}=1$ (i.e., $\hat{A}^\beta = e^{-\beta\hat{H}}$, the Boltzmann operator itself) is not invariant with the approximate evolution of the LSC-IVR for anharmonic systems[33, 64, 92, 126, 127, 131-134]. The latter may lead to the unphysical decay (i.e., overdamping) intrinsic in the LSC-IVR correlation function, which becomes progressively worse for longer times[33, 135]. As a part of the effects due to the failure of the LSC-IVR to conserve the quantum canonical distribution, the LSC-IVR reaction rate depends on the choice of the dividing surface[58, 128]. (It is sometimes not necessarily a drawback since this dependence may be used variationally to identify the dividing surface.) As another part of the effects, the "artificial energy flow" exists from intramolecular to intermolecular modes (also denoted "zero point energy leakage") for such a complex polyatomic system as liquid water [e.g., see the note (Ref. 36) of Ref. [136], Ref. [28], Ref. [133], Ref. [33], etc.]. The accuracy of the LSC-IVR correlation function then depends on the competition between the unphysical decay caused by the method and the inherent physical decay of the "real" system[33]. Because the Raman spectroscopy of liquid water involves ultra-fast dynamics [as will also be demonstrated in Fig. 2 the physical time scale of the (relevant) polarizability-derivative correlation function is relatively short—about 2~3 hundred femtoseconds], the intrinsic unphysical decay of the LSC-IVR correlation function will be greatly compensated by the inherent physical decay in the system. This is actually the main reason why the LSC-IVR is a good approximate quantum approach for study of the vibrational spectroscopy in condensed phase systems[33, 59].

Regardless of these two drawbacks, the LSC-IVR describes various aspects of the (short-time) dynamics very well[33, 52, 56-59, 129-131, 134, 137-147]. For instance, the LSC-IVR has been shown to describe reactive flux correlation functions for chemical reaction rates quite well (when the dividing surface is chosen in the reasonable region), including strong tunneling regimes[53, 58, 148], and correlation functions for vibrational spectra, transport properties, vibrational/electronic energy transfer[28, 33, 56, 59, 129-131, 134, 138-147] in systems with enough degrees of freedom for true quantum coherence effects to be unimportant and where the physical decay time is relatively short. When the vibrational spectroscopy is studied, the LSC-IVR is exact for any long time in the harmonic limit, regardless of whether the involved operator is a linear or nonlinear function of coordinates or/and momenta and independent of whether it is in the low frequency regime or involves high frequency vibrations. The LSC-IVR maintains all aspects of the classical coherence and does not suffer the "curvature problem[94]" or "artificial frequencies and resonances[27, 94, 149]" in the simulation.

**B. Local Gaussian approximation**

The Wigner function for operator $\hat{B}$ in Eq. (11) is often trivial to obtain. When $\hat{B}$ is a function only of coordinates or only of momenta, its Wigner function is simply the classical function itself. In contrast, calculation of the Wigner function for operator $\hat{A}^\beta$ is non-trivial, because it involves the Boltzmann operator with the total Hamiltonian of the complete system and the multidimensional Fourier transform. It is also important to do this in order to yield the distribution of initial conditions of momenta $\mathbf{p}_0$ before propagating the real time trajectories. To accomplish this task, several approximations[53, 56, 138, 141, 147] were introduced for the LSC-IVR. Superior to all these approximations, the local Gaussian approximation (LGA) for treating imaginary frequencies[58] with the LSC-IVR provides an *indeed* practical tool to study quantum effects in general large/complex molecular systems whose interactions are often too difficult to be parameterized by Gaussians or polynomials and where *ab initio* calculations are called for. Because it is straightforward to implement the LGA with state-of-art PIMD methods (including ring polymer contraction[45], color noise methods[46, 47], and the more generally applicable "middle" thermostat scheme[48-51]), the LGA can prepare the initial conditions for the LSC-IVR both efficiently and reasonably accurately. Below we briefly summarize the LGA, which is what we use for the study of the Raman spectroscopy of liquid water in this paper.

The Hamiltonian around $\mathbf{x}$ can be expanded to 2nd order as

$$H(\mathbf{x}+\Delta\mathbf{x}) \approx \frac{1}{2}\mathbf{p}^T \mathbf{M}^{-1}\mathbf{p} + V(\mathbf{x}) + \left(\frac{\partial V}{\partial \mathbf{x}}\right)^T \Delta\mathbf{x} + \frac{1}{2}\Delta\mathbf{x}^T \frac{\partial^2 V}{\partial \mathbf{x}^2}\Delta\mathbf{x} \qquad (13)$$

As in the standard normal-mode analysis, mass-weighted Hessian matrix elements are given by

$$\mathcal{H}_{kl} = \frac{1}{\sqrt{m_k m_l}} \frac{\partial^2 V}{\partial x_k \partial x_l} \quad (14)$$

where $m_k$ represents the mass of the $k$-th degree of freedom with $3N$ the total number of degrees of freedom. The eigenvalues of the mass-weighted Hessian matrix produce normal-mode frequencies $\{\omega_k\}$, i.e.,

$$\mathbf{T}^T \mathcal{H} \mathbf{T} = \boldsymbol{\lambda} \quad (15)$$

with $\boldsymbol{\lambda}$ a diagonal matrix with the elements $\{(\omega_k)^2\}$ and $\mathbf{T}$ an orthogonal matrix. If $\mathbf{M}$ is the diagonal 'mass matrix' with elements $\{m_k\}$, then the mass-weighted normal mode coordinates and momenta $(\mathbf{X}, \mathbf{P})$ are given in terms of the Cartesian variables $(\mathbf{x}, \mathbf{p})$ by

$$\mathbf{X} = \mathbf{T}^T \mathbf{M}^{1/2} \mathbf{x} \quad , \quad (16)$$

$$\mathbf{P} = \mathbf{T}^T \mathbf{M}^{-1/2} \mathbf{p} \quad , \quad (17)$$

and

$$\Delta \mathbf{X} = \mathbf{T}^T \mathbf{M}^{1/2} \Delta \mathbf{x} \quad . \quad (18)$$

Eq. (13) can be expressed as

$$H(\mathbf{x}+\Delta\mathbf{x}) \equiv H(\mathbf{X}+\Delta\mathbf{X}) \approx \frac{1}{2}\mathbf{P}^T\mathbf{P} + V(\mathbf{X}) + \Delta\mathbf{X}^T\mathbf{F} + \frac{1}{2}\Delta\mathbf{X}^T\boldsymbol{\lambda}\,\Delta\mathbf{X} \quad (19)$$

with $\mathbf{F}$ as the first-derivative in the mass-weighted normal mode coordinates

$$\mathbf{F} = \mathbf{T}^T\mathbf{M}^{-1/2}\left(\frac{\partial V}{\partial \mathbf{x}}\right). \quad (20)$$

By virtue of the fact that

$$\frac{\left\langle x - \frac{\Delta x}{2} \middle| e^{-\beta \hat{H}} \middle| x + \frac{\Delta x}{2}\right\rangle}{\left\langle x \middle| e^{-\beta \hat{H}} \middle| x\right\rangle} = \exp\left[-\frac{mQ(u)}{2\hbar^2\beta}(\Delta x)^2\right] \quad (21)$$

for the 1-dimensional harmonic case which was implemented in LHA by Shi and Geva[141], it is straightforward to show the Wigner function of the Boltzmann operator $e^{-\beta \hat{H}}$ is given by

$$\mathcal{P}_W^{eq}(\mathbf{x},\mathbf{P}) = \left\langle \mathbf{x}\middle| e^{-\beta\hat{H}}\middle|\mathbf{x}\right\rangle \prod_{k=1}^{3N}\left[\left(\frac{\beta}{2\pi Q(u_k)}\right)^{1/2} \exp\left[-\beta\frac{(P_k)^2}{2Q(u_k)}\right]\right], \quad (22)$$

where $u_k = \beta\hbar\omega_k$, $P_k$ is the $k$-th component of the mass-weighted normal-mode momenta $\mathbf{P}$ (in Eq. (17)) and the quantum correction factor with the LGA ansatz proposed by Liu and Miller[58] for both real and imaginary frequencies is given by

$$Q(u) = \begin{cases} \dfrac{u/2}{\tanh(u/2)} & \text{for real } u \\[6pt] \dfrac{1}{Q(u_i)} = \dfrac{\tanh(u_i/2)}{u_i/2} & \text{for imaginary } u\ (u = iu_i) \end{cases}. \quad (23).$$

In terms of the phase space variables $(\mathbf{x}, \mathbf{p})$, Eq. (22) thus becomes

$$\mathcal{P}_W^{\text{eq, LGA}}(\mathbf{x}, \mathbf{p}) = \langle \mathbf{x} | e^{-\beta \hat{H}} | \mathbf{x} \rangle \left( \frac{\beta}{2\pi} \right)^{3N/2} \left| \det \left( \mathbf{M}_{\text{therm}}^{-1}(\mathbf{x}) \right) \right|^{1/2} \exp \left[ -\frac{\beta}{2} \mathbf{p}^T \mathbf{M}_{\text{therm}}^{-1} \mathbf{p} \right] \quad (24)$$

with the thermal mass matrix $\mathbf{M}_{\text{therm}}$ given by

$$\mathbf{M}_{\text{therm}}^{-1}(\mathbf{x}) = \mathbf{M}^{-1/2} \mathbf{T} \mathbf{Q}(\mathbf{u})^{-1} \mathbf{T}^T \mathbf{M}^{-1/2} \quad (25)$$

and the diagonal matrix $\mathbf{Q}(\mathbf{u}) = \text{diag}\{Q(u_k)\}$.

The explicit form of the LSC-IVR correlation function [Eq. (9)] with the LGA is thus given by

$$C_{AB}^{\text{LSC-IVR}}(t) = \frac{1}{Z} \int d\mathbf{x}_0 \langle \mathbf{x}_0 | e^{-\beta \hat{H}} | \mathbf{x}_0 \rangle \int d\mathbf{P}_0 \prod_{k=1}^{3N} \left[ \left( \frac{\beta}{2\pi Q(u_k)} \right)^{1/2} \exp \left[ -\beta \frac{(P_{0,k})^2}{2Q(u_k)} \right] \right] \\ \times f_A(\mathbf{x}_0, \mathbf{p}_0) B(\mathbf{x}_t, \mathbf{p}_t) \quad (26)$$

where

$$f_A(\mathbf{x}_0, \mathbf{p}_0) = \int d\Delta\mathbf{x} \frac{\langle \mathbf{x}_0 - \frac{\Delta\mathbf{x}}{2} | \hat{A}^\beta | \mathbf{x}_0 + \frac{\Delta\mathbf{x}}{2} \rangle}{\langle \mathbf{x}_0 | e^{-\beta \hat{H}} | \mathbf{x}_0 \rangle} e^{i\Delta\mathbf{x}^T \cdot \mathbf{p}_0 / \hbar} \bigg/ \int d\Delta\mathbf{x} \frac{\langle \mathbf{x}_0 - \frac{\Delta\mathbf{x}}{2} | e^{-\beta \hat{H}} | \mathbf{x}_0 + \frac{\Delta\mathbf{x}}{2} \rangle}{\langle \mathbf{x}_0 | e^{-\beta \hat{H}} | \mathbf{x}_0 \rangle} e^{i\Delta\mathbf{x}^T \cdot \mathbf{p}_0 / \hbar} \quad (27)$$

is a function depending on the operator $\hat{A}^\beta$. The procedure for implementing the LSC-IVR (LGA) [58] is described as follows:

(1) Use PIMD to simulate the system in thermal equilibrium. In this paper we employ the generally applicable "middle" thermostat scheme for PIMD[48, 49]. It improves over the conventional PIMD algorithms on the sampling efficiency of the path integral beads by about an order of magnitude for general systems. It should be stressed that the "middle" thermostat scheme can be combined with other techniques (ring polymer contraction[45], color noise methods[46, 47]) to gain even more efficiency.

(2) At specific time steps in PIMD, randomly select one path integral bead as the initial configuration $\mathbf{x}_0$ for the real time dynamics. Diagonalize the mass-weighted Hessian matrix of the potential surface to obtain the local normal mode frequencies [i.e., Eq. (15)].

(3) The LGA produces the Gaussian distribution for mass-weighted normal mode momenta

$\prod_{k=1}^{3N} (\beta / 2\pi Q(u_k))^{1/2} \exp\left[ -\beta (P_{0,k})^2 / (2Q(u_k)) \right]$, which is used to sample the initial Cartesian momenta $\mathbf{p}_0 = \mathbf{M}^{1/2} \mathbf{T} \mathbf{P}_0$ for a real time trajectory.

(4) Run the real time classical trajectory (with constant energy) from the phase space point $(\mathbf{x}_0, \mathbf{p}_0)$ and estimate the property $f_A(\mathbf{x}_0, \mathbf{p}_0) B(\mathbf{x}_t(\mathbf{x}_0, \mathbf{p}_0), \mathbf{p}_t(\mathbf{x}_0, \mathbf{p}_0))$ for the corresponding time correlation function.

(5) Repeat steps (2)-(4) and sum the property $f_A(\mathbf{x}_0, \mathbf{p}_0) B(\mathbf{x}_t(\mathbf{x}_0, \mathbf{p}_0), \mathbf{p}_t(\mathbf{x}_0, \mathbf{p}_0))$ for all real time classical trajectories until a converged result is reached.

It should be stressed that *no* approximation for the potential energy surface is made in Step (1) (sampling the beads) and Step (4) (the real time dynamics of trajectories).

## III. Raman spectrum and polarizability-derivative correlation function

It is straightforward to verify the relation

$$\hat{A}^{\beta}_{\text{Kubo}} = \frac{i}{\beta\hbar}\left[\hat{A}, e^{-\beta\hat{H}}\right] \tag{28}$$

and the Taylor expansion

$$A(\mathbf{x}+\Delta\mathbf{x}) = A(\mathbf{x}) + \frac{\partial A(\mathbf{x})}{\partial \mathbf{x}}\cdot\Delta\mathbf{x} + \frac{1}{2}\Delta\mathbf{x}^T \cdot \frac{\partial^2 A(\mathbf{x})}{\partial \mathbf{x}^2}\cdot\Delta\mathbf{x} + O(\Delta\mathbf{x}^3) \ . \tag{29}$$

For the LSC-IVR (LGA) formulation [Eq. (26)], one can obtain

$$f_A(\mathbf{x}_0,\mathbf{p}_0)B(\mathbf{x}_t,\mathbf{p}_t) \approx \left[\left(\frac{\partial \alpha_{\text{iso}}}{\partial \mathbf{x}_0}\right)^T \mathbf{M}^{-1}_{\text{therm}}(\mathbf{x}_0)\mathbf{p}_0\right]\dot{\alpha}_{\text{iso}}(\mathbf{x}_t,\mathbf{p}_t) \tag{30}$$

for $\left\langle \hat{\alpha}_{\text{iso}}(0)\hat{\dot{\alpha}}_{\text{iso}}(t)\right\rangle_{\text{Kubo}}$ and

$$f_A(\mathbf{x}_0,\mathbf{p}_0)B(\mathbf{x}_t,\mathbf{p}_t) \approx \text{Tr}\left\{\left[\left(\frac{\partial \boldsymbol{\beta}}{\partial \mathbf{x}_0}\right)^T \mathbf{M}^{-1}_{\text{therm}}(\mathbf{x}_0)\mathbf{p}_0\right]\dot{\boldsymbol{\beta}}(\mathbf{x}_t,\mathbf{p}_t)\right\} \tag{31}$$

for $\left\langle \text{Tr}\left[\hat{\boldsymbol{\beta}}(0)\hat{\dot{\boldsymbol{\beta}}}(t)\right]\right\rangle_{\text{Kubo}}$. In Eq. (31) $(\partial\boldsymbol{\beta}/\partial\mathbf{x}_0)^T\mathbf{M}^{-1}_{\text{therm}}(\mathbf{x}_0)\mathbf{p}_0$ is a $3\times 3$ matrix whose elements are

$$\left[\left(\frac{\partial \boldsymbol{\beta}}{\partial \mathbf{x}_0}\right)^T \mathbf{M}^{-1}_{\text{therm}}(\mathbf{x}_0)\mathbf{p}_0\right]_{ij} = \left(\frac{\partial \beta_{ij}}{\partial \mathbf{x}_0}\right)^T \mathbf{M}^{-1}_{\text{therm}}(\mathbf{x}_0)\mathbf{p}_0 \quad (i,j\in\{x,y,z\}) \ . \tag{32}$$

Note that the collective polarizability $\boldsymbol{\alpha}$ may be so complicated that the analytical form for its derivative is too tedious to obtain. On the other hand, it involves much effort to use the finite difference to numerically evaluate the derivative over the coordinate $\partial \mathbf{\alpha}/\partial \mathbf{x}_0$ when the total number of the degrees of freedom of the system is large. It is thus difficult to directly calculate either $\left( \dfrac{\partial \alpha_{iso}}{\partial \mathbf{x}_0} \right)^T \mathbf{M}_{therm}^{-1}(\mathbf{x}_0)\mathbf{p}_0$ in Eq. (30) or $\left[ \left( \dfrac{\partial \mathbf{\beta}}{\partial \mathbf{x}_0} \right)^T \mathbf{M}_{therm}^{-1}(\mathbf{x}_0)\mathbf{p}_0 \right]$ in Eq. (31). To efficiently estimate the term we introduce a numerical trick by using the equality

$$\left( \frac{\partial g}{\partial \mathbf{x}_0} \right)^T \mathbf{M}_{therm}^{-1}\mathbf{p}_0 = \lim_{\varepsilon \to 0} \frac{g\left(\mathbf{x}_0 + \varepsilon \mathbf{M}_{therm}^{-1}\mathbf{p}_0\right) - g\left(\mathbf{x}_0 - \varepsilon \mathbf{M}_{therm}^{-1}\mathbf{p}_0\right)}{2\varepsilon} \quad . \tag{33}$$

Here, $g$ is a coordinate-dependent function (e.g., $\mathbf{\alpha}$, $\alpha_{iso}$ or $\mathbf{\beta}$). Eq. (33) suggests that a good estimator for $\left( \dfrac{\partial g}{\partial \mathbf{x}_0} \right)^T \mathbf{M}_{therm}^{-1}\mathbf{p}_0$ is offered by the finite difference along the direction of the vector $\mathbf{M}_{therm}^{-1}(\mathbf{x}_0)\mathbf{p}_0$, i.e.,

$$\left( \frac{\partial g}{\partial \mathbf{x}_0} \right)^T \mathbf{M}_{therm}^{-1}\mathbf{p}_0 \approx \frac{g\left(\mathbf{x}_0 + \varepsilon \mathbf{M}_{therm}^{-1}\mathbf{p}_0\right) - g\left(\mathbf{x}_0 - \varepsilon \mathbf{M}_{therm}^{-1}\mathbf{p}_0\right)}{2\varepsilon} \quad , \tag{34}$$

where $\varepsilon$ represents a small enough (positive) number to guarantee the numerical convergence. $\varepsilon$ should be adjusted as the vector $\mathbf{M}_{therm}^{-1}\mathbf{p}_0$ varies. A reasonable choice of $\varepsilon$ could be

$$\varepsilon = \frac{\delta x}{\left\| \mathbf{M}_{therm}^{-1}\mathbf{p}_0 \right\|} \quad , \tag{35}$$

where $\delta x$ is a (positive) constant parameter dependent on the system and $\left\| \mathbf{M}_{therm}^{-1}\mathbf{p}_0 \right\|$ represents the length (i.e., $L^2$ norm) of the vector $\mathbf{M}_{therm}^{-1}\mathbf{p}_0$.

Similarly, one has

$$f_A(\mathbf{x}_0,\mathbf{p}_0)B(\mathbf{x}_t,\mathbf{p}_t) \approx \dot{\alpha}_{iso}(\mathbf{x}_0,\mathbf{p}_0)\dot{\alpha}_{iso}(\mathbf{x}_t,\mathbf{p}_t) \tag{36}$$

for the real part of $\left\langle \hat{\dot{\alpha}}_{iso}(0)\hat{\dot{\alpha}}_{iso}(t) \right\rangle_{std}$ and

$$f_A(\mathbf{x}_0,\mathbf{p}_0)B(\mathbf{x}_t,\mathbf{p}_t) \approx \text{Tr}\left[\dot{\boldsymbol{\beta}}(\mathbf{x}_0,\mathbf{p}_0)\dot{\boldsymbol{\beta}}(\mathbf{x}_t,\mathbf{p}_t)\right] \tag{37}$$

for the real part of $\left\langle \text{Tr}\left[\hat{\dot{\boldsymbol{\beta}}}(0)\hat{\dot{\boldsymbol{\beta}}}(t)\right] \right\rangle_{std}$. It is easy to verify that Eqs. (30)-(31) involve $O(\Delta\mathbf{x}^3)$ in accuracy of Eq. (29) while Eqs. (36)-(37) involve only $O(\Delta\mathbf{x}^2)$ in accuracy of Eq. (29). We then use Kubo-transformed correlation functions in the simulations in the paper. That is, Eq. (26) with Eqs. (30)-(31) for the LSC-IVR correlation functions and then Eqs. (7)-(8) for the isotropic and anisotropic Raman spectra.

## IV. Results and discussions

### A. Simulation details

The LSC-IVR simulation was carried out at $T = 298.15$ K for both liquid water and heavy water. The density of liquid water and that of heavy water are $0.997 \text{ g}\cdot\text{cm}^{-3}$ and $1.1045 \text{ g}\cdot\text{cm}^{-3}$, respectively. The system was simulated with 216 $H_2O$ or $D_2O$ molecules in a box with periodic boundary conditions. The charge-charge, charge-dipole and dipole-dipole interactions were treated using Ewald summation, while the repulsion-dispersion forces and the electrostatic interactions including quadrupoles were cut off from 9.0 Å to 9.1 Å with the smoothing function described in Appendix of Ref. [115]. Staging PIMD was performed with $N_P = 48$ beads for the NVT ensemble. The Langevin thermostat in the "middle" scheme[48-50] with a time step of $\Delta t = 0.1\,\text{fs}$ (the time step size $\Delta t = 0.7\,\text{fs}$ is already small enough to obtain fully converged PIMD results with 48 beads) was implemented for propagating the PIMD trajectories. The initial configurations of the classical trajectories were obtained by randomly selecting one of the path integral beads every 0.1 ps in the imaginary time propagation of PIMD. Each classical trajectory in the LSC-IVR was propagated up to 1 ps. For the real time dynamics in the LSC-IVR, the velocity-Verlet algorithm was employed with a time step of $\Delta t = 0.1\,\text{fs}$. This small time step guarantees the accuracy of the polarizability-derivative $\dot{\boldsymbol{\alpha}}(t)$ by using the finite difference $[\boldsymbol{\alpha}(t+\Delta t) - \boldsymbol{\alpha}(t-\Delta t)]/2\Delta t$ along the real time trajectory. We use $\delta x = 10^{-5}\,\text{Å}$ for the finite difference in Eq. (35) for obtaining the converged result for the Kubo-transformed correlation function. In each simulation 12000 such trajectories were used for evaluating the correlation functions. (24000 trajectories were used for the LSC-IVR results in Figs. 2-4.)

Similarly, the classical correlation functions were obtained by propagating 32 classical (constant energy) trajectories with the initial conditions sampled from the classical NVT ensemble. Each trajectory was propagated up to 250 ps. Time averaging was employed for computing the correlation functions along the classical NVE trajectory.

When O-H bond lengths or H-O-H bond angles were constrained, we implemented the RATTLE algorithm[150, 151] in PIMD and in the LSC-IVR for quantum simulations and did so in MD for classical simulations. When the two O-H bond lengths of each water molecule were constrained, the values were fixed at 0.917 Å—the optimal O-H bond length for a water monomer. When the H-O-H bond angle of each water molecule was constrained, its value was fixed at 104.508°—the optimal H-O-H bond angle for a water monomer.

**B. Results and discussions**

**1. Raman spectra of liquid water at 298.15 K**

We first investigate the isotropic and anisotropic Raman spectra of liquid water at 298.15 K using the POLI2VS water force field. Fig. 1 compares the density distribution of local normal-mode frequencies of quantum configurations sampled from PIMD to that obtained from the classical canonical ensemble. Even at room temperature, about 16% of local normal-mode frequencies generated by PIMD are imaginary, with more than 6% in the "deep tunneling" imaginary frequency regime ($\beta\hbar|\omega| \geq \pi$). In contrast, about 12% of local frequencies generated by classical MD are imaginary, with only 0.7% access the "deep tunneling" imaginary frequency regime. Fig. 1 demonstrates that the quantum and classical results are close to each other in the hindered motion region (0-400 cm$^{-1}$). It is then expected that quantum Raman spectra are not much different from classical results in this low-frequency region (0-400 cm$^{-1}$). Interestingly, considerable quantum effects appear in from the high-frequency region of the librational motion to the intermediate regime between the librational and bending bands, which arises from the increase of the librational frequency caused by the elongation of O-H bonds in PIMD configurations. The quantum density is slightly lower than the classical one in the bending region (~1750 cm$^{-1}$). Fig. 1 demonstrates that the classical density almost vanishes but the quantum density is much higher in the intermediate region between the bending and stretching bands (2000-2600 cm$^{-1}$). This hints that significant nuclear quantum effects are involved in this region and that classical MD may miss important peaks in the experimental spectrum. The quantum density of stretching frequencies is broad, while the classical result is much narrower and blue-shifted. We then expect that the stretching band in the classical Raman spectra may be considerably narrower and blue-shifted from that produced by the quantum simulation or that measured in experiment.

Fig. 2a shows the comparison of the LSC-IVR polarizability-derivative correlation function to the classical one for the isotropic Raman spectrum, while Fig. 2b does so for the anisotropic Raman spectrum. In either panel of Fig. 2 the LSC-IVR predicts a more rapid decay of the amplitude and longer oscillation periods of the relevant polarizability-derivative correlation function than is suggested by the MD simulation. The anisotropic polarizability-derivative correlation function decays slightly faster than the isotropic one. The former (Fig. 2b) decays to zero within 200 fs and the latter (Fig. 2a) does so within 250 fs. This indicates that the Raman spectroscopy of liquid water is connected to ultra-fast dynamics, where the LSC-IVR describes quantum dynamics reasonably well.

Fig. 3 shows the LSC-IVR isotropic Raman spectrum of liquid water at 298.15 K in comparison to the classical result as well as to the experimental data reported in Ref. [76]. Fig. 4 does so for the anisotropic

Raman spectrum. In either of Fig. 3 and Fig. 4 the intensity is normalized to its maximum value. (This is also the case for any other spectra throughout the paper.) We first consider the frequency region below 1000 cm$^{-1}$. Although the LSC-IVR and MD show a qualitatively correct picture in the librational region (below 1000 cm$^{-1}$) of the isotropic Raman spectrum (Fig. 3a), they demonstrate poor results in the same region of the anisotropic Raman spectrum (Fig. 4a). When Hasegawa and Tanimura developed the POLI2VS force field, charge fluctuations between the water molecules were not implemented in the water model, which leads to an inaccurate description for low-frequency Raman modes[112, 115]. This can be consistently improved by taking care of intramolecular charge flow (CF) effects, intermolecular charge transfer (CT) effects, and intermolecular dipole-induced-dipole (DID) effects in the polarizability function of the POLI2VS force field[112]. The revPBE0-D3 DFT functional[107] or the MB-pol water force field[103, 105] seems to work reasonably well in the low frequency region[40, 42]. It will be interesting in future to use the LSC-IVR with an improved version of the POLI2VS, the revPBE0-D3 functional, or the MB-pol water force field to study low frequency Raman modes for liquid water.

We then focus on the frequency region that is greater than 1000 cm$^{-1}$, where the POLI2VS force field is expected to be reasonable. First consider the Raman modes between 1000 cm$^{-1}$ and 2500 cm$^{-1}$, where much information is available. The LSC-IVR shows more satisfactory agreement with experiment than MD for the H-O-H bending band. The peak position of the H-O-H bending band produced by the LSC-IVR is blue-shifted by ~62 cm$^{-1}$ from experiment in the isotropic Raman spectrum and by ~71 cm$^{-1}$ in the anisotropic Raman spectrum, while that for MD is blue-shifted by ~87 cm$^{-1}$ in the isotropic spectrum and by ~107 cm$^{-1}$ in the anisotropic spectrum (as shown in Figs. 3a and 4a). Both the relative intensity and the structure of the spectrum are better reproduced by the LSC-IVR than by MD in the intermediate region between the librational and bending bands (1000-1500 cm$^{-1}$) and also in the intermediate region between the bending and stretching bands (1800-2500 cm$^{-1}$). For instance, Fig. 3a shows that in the isotropic Raman spectrum the shoulder around 1351 cm$^{-1}$ and the peak around 2061 cm$^{-1}$ are well captured by the LSC-IVR but completely absent in the classical MD results. This suggests that significant nuclear quantum effects exist in these intermediate regions, which is consistent with our previous study on liquid water[33]. The peak around 2061 cm$^{-1}$ in the isotropic Raman spectrum (in Fig. 3a) was usually assigned to a combination of the bending-mode and the librational-mode, the counterpart of which is around 2110 cm$^{-1}$ in the anisotropic Raman spectrum (in Fig. 4a). We will shortly examine whether the conventional assignment is reasonable by selectively freezing the bending or stretching mode in the simulation. The LSC-IVR with the POLI2VS water force field is able to semi-quantitatively describe the Raman peak. For comparison, CMD with the revPBE0-D3 functional[107] is also able to capture the Raman peak[42], but CMD with the MB-pol water force field[103, 105] completely misses it in the anisotropic Raman spectrum[40]. It will be interesting to systematically investigate whether other quantum dynamics methods with the MB-pol water force field[103, 105] faithfully describes the combination of the bending-mode and the librational-mode in the Raman spectroscopy of liquid water.

We then study the Raman modes in the high frequency region—the O-H stretching band. The peak position of the O-H stretching band produced by the LSC-IVR is blue-shifted by ~25 cm$^{-1}$ from the experimental result in the isotropic Raman spectrum (Fig. 3b) and by ~74 cm$^{-1}$ in the anisotropic Raman spectrum (Fig. 4b), while that yielded by classical MD is blue-shifted by ~153 cm$^{-1}$ from experiment in the isotropic Raman spectrum (Fig. 3b) and by ~174 cm$^{-1}$ in the anisotropic Raman spectrum (Fig. 4b). In experiment the full width at half maximum (FWHM) of the O-H stretching band is ~416 cm$^{-1}$ in the isotropic Raman spectrum (Fig. 3b) and ~325 cm$^{-1}$ in the anisotropic Raman spectrum (Fig. 4b). The LSC-IVR agrees well with experiment—the LSC-IVR FWHM is ~415 cm$^{-1}$ in the isotropic Raman spectrum (Fig. 3b) and ~342 cm$^{-1}$ in the anisotropic Raman spectrum (Figs. 4b and 4c). In contrast, the FWHM produced by classical MD is ~269 cm$^{-1}$ in the isotropic Raman spectrum (Fig. 3b) and ~247 cm$^{-1}$ in the anisotropic Raman spectrum (Figs. 4b and 4c), which is significantly narrower than the corresponding experimental value.

It is encouraging to see in Fig. 3b that the LSC-IVR reproduces two peaks in the O-H stretching band in the experimental isotropic Raman spectrum. The peak at ~3252 cm$^{-1}$ has long been suspected to be contributed from the bending-mode overtone and its Fermi resonance with the stretching-mode, which has a relatively higher intensity than the peak of stretching-mode (~3384 cm$^{-1}$) at room temperature. In contrast, the classical MD simulation only demonstrates one peak for room-temperature liquid water, which is not even qualitatively correct. This indicates that quantum dynamical effects are decisive in the O-H stretching band of the isotropic Raman spectrum. The feature in the experimental isotropic Raman spectrum, however, is captured by neither CMD nor RPMD simulations[42] with the revPBE0-D3 functional[107]. Such a feature is also not distinct in the MD or CMD isotropic Raman spectrum (for room-temperature liquid water)[152] with the MB-pol water force field[103, 105].

To better understand the Raman modes in the frequency region 1000-4000 cm$^{-1}$ of liquid water, we propose two types of computer "experiments". One is to freeze the intramolecular O-H stretching mode, the other is to freeze the intramolecular H-O-H bending mode. The two types of computer experiments will provide insights on the interplay of the O-H stretching mode (or the H-O-H bending mode) and other intra- and inter-molecular modes.

2. **Raman spectra of liquid water with constrained intramolecular O-H bond lengths**

When the O-H bond length of each water molecule is fixed at its optimal value for a monomer in the simulation, Panel (a) of Fig. 5 compares the normalized classical and LSC-IVR isotropic polarizability-derivative auto-correlation functions, and Panel (b) of Fig. 5 does so for the anisotropic polarizability-derivative auto-correlation functions. Their Fourier transforms lead to the isotropic Raman spectrum in Fig. 6 and the anisotropic one in Fig. 7. (In both figures the experimental results and the Raman spectra yielded without any constraints are plotted for comparison.) As expected to be a consequence of the constrained dynamics, the stretching band vanishes in both the isotropic and anisotropic Raman spectra. When the O-H bond length is fixed, the peak of the bending band is blue-shifted by ~48 cm$^{-1}$ in the isotropic Raman spectrum and by ~34 cm$^{-1}$ in the anisotropic Raman spectrum. This suggests that the Raman excitation energy in the bending band of liquid water is lowered by the coupling between the stretching mode and the bending mode.

Interestingly, the intensity of the intermediate region (2000-2400 cm$^{-1}$) between the stretching and bending bands almost disappears in the isotropic Raman spectrum (as shown in Fig. 6a), which indicates that the peak of this intermediate region in the isotropic Raman spectrum is mostly caused by the interaction between the stretching motion and other types of motions. In contrast, the intensity of intermediate region (2000-2400 cm$^{-1}$) is greatly diminished but the peak is still noticeable in the anisotropic Raman spectrum (as shown in Fig. 7a). This suggests that a substantial part of the signal in the intermediate region (2000-2400 cm$^{-1}$) in the anisotropic Raman spectrum involves contributions from other combined motions that are independent of the stretching mode, for instance, from the combination of the bending mode and the librational mode. We will verify this by freezing the bending motion.

As shown in Figs. 6-7, when the stretching motion is frozen, the intensity in the low-frequency region (below 1000 cm$^{-1}$) is significantly weakened in the isotropic Raman spectrum but does not change much in the anisotropic Raman spectrum. Because the POLI2VS force field does not describe this low-frequency region consistently well, it will be necessary to investigate this with other accurate water force fields or electronic structure methods to obtain more insights.

Comparisons of the corresponding classical spectra in Figs. 6c, 6d, 7c and 7d lead to similar observations. The intensity in the intermediated region (2000-2400 cm$^{-1}$) between the bending and stretching bands in the classical isotropic Raman spectrum mainly arises from the interplay of the stretching motion and other types of motions. Classical MD also predicts that a considerable portion of the peak of this intermediated region (2000-2400 cm$^{-1}$) in the anisotropic Raman spectrum should be irrelevant to the stretching motion.

**3. Raman spectra of liquid water with constrained intramolecular H-O-H bond angles**

We then fix the H-O-H bond angle of each water molecule at its optimal value for a monomer in the simulation of liquid water. When such a type of constrained dynamics is applied, Panel (a) [or (b)] of Fig. 8 compares the normalized classical and LSC-IVR isotropic (or anisotropic) polarizability-derivative auto-correlation functions. Their Fourier transforms produce the isotropic Raman spectrum in Fig. 9 and the anisotropic one in Fig. 10. In both figures the experimental data and the Raman spectra generated without any constraints are also shown for comparison. As expected when the bending motion is frozen, the bending peak in the isotropic Raman spectrum (~1625 cm$^{-1}$) disappears, so does the bending peak in the anisotropic Raman spectrum (~1644 cm$^{-1}$).

Fig. 9a and Fig. 10a show that the peak in the intermediate region between the bending and stretching bands almost vanishes in the isotropic and anisotropic Raman spectra when we freeze the bending mode. It may then be concluded from Fig. 6a and Fig. 9a that the peak in this intermediate region in the isotropic

Raman spectrum mainly arises from the interplay of the stretching motion and the bending motion. In contrast, Fig. 7a and Fig. 10a suggest that a substantial part of the peak in this intermediate region of the anisotropic Raman spectrum should be attributed to the combined motion of the bending mode and the librational mode.

Fig. 9b demonstrates that only one peak appears in the stretching band of the isotropic Raman spectrum when the bending mode is frozen. For comparison two peaks arise when no constraints are applied. This indicates the attribution of the peak at 3230~3260 cm$^{-1}$ in the isotropic Raman spectrum to the bending mode overtone and its Fermi resonance with the stretching mode. In contrast to the isotropic Raman spectrum, the anisotropic Raman spectrum (in Fig. 10b) displays little changes when we freeze the bending mode. This hints that the bending motion has little effect on the stretching band of the anisotropic Raman spectrum.

When the H-O-H bending motion is frozen, the intensity in the librational region (below 1000 cm$^{-1}$) is reduced in the isotropic Raman spectrum (Fig. 9a) but does not change much in the anisotropic Raman spectrum (Fig. 10a). This is similar to the variations in the two types of Raman spectra when the O-H bond length is fixed. Figs. 6a, 7a, 9a, and 10a then indicate that in the isotropic Raman spectrum the low-frequency region involves considerable coupling effects between intermolecular modes and intramolecular (bending and stretching) modes, while such coupling effects between intermolecular and intramolecular motions are much less noticeable in the low-frequency region of the anisotropic Raman spectrum.

Similarly, Figs. 9c and 9d compares the original classical Raman isotropic spectrum to the one with constrained intramolecular H-O-H angles, while Figs. 10c and 10d does so for the classical Raman anisotropic spectrum. Classical MD also suggests that the bending motion has little effect on the stretching band of the anisotropic Raman spectrum but significantly influences the lower frequency region of the stretching band of the isotropic Raman spectrum. One may see from Fig. 7c and Fig. 10c that a considerable portion of the peak in the intermediated region (2000-2400 cm$^{-1}$) between the bending and stretching bands in the classical anisotropic Raman spectrum is also attributed to the combination of the bending and librational modes.

## 4. Raman spectra of heavy water at 298.15 K

To investigate isotope effects in the Raman spectroscopy we study pure liquid heavy water at 298.15 K using the POLI2VS water force field. Fig. 11 shows the density distribution of local normal-mode frequencies of quantum configurations sampled from PIMD. The classical density distribution is plotted for comparison. The comparison in Fig. 11 for heavy water is similar to that in Fig. 1 for liquid water. In heavy water at room temperature, about 15% of local frequencies yielded by PIMD are imaginary, with more than 6% in the "deep tunneling" imaginary frequency regime ($\beta\hbar|\omega| \geq \pi$). In contrast, about 12% of local frequencies produced by classical MD are imaginary, with only 0.06% access the "deep tunneling" imaginary frequency regime.

Fig. 12a compares the LSC-IVR polarizability-derivative correlation function to the classical one for the isotropic Raman spectrum of heavy water, while Fig. 12b does so for the anisotropic Raman spectrum. Similar to Fig. 2, in either panel of Fig. 12 the LSC-IVR predicts a more rapid decay of the amplitude and longer oscillation periods of the relevant polarizability-derivative correlation function than is suggested by the classical MD simulation. The anisotropic polarizability-derivative correlation function (Fig. 12b) decays to zero within 250 fs and the isotropic one (Fig. 12a) does so within 300 fs, either of which for heavy water decays less rapidly than its counterpart for liquid water.

Fig. 13 compares the LSC-IVR isotropic Raman spectrum of heavy water at 298.15 K to the classical result as well as to the experimental data[76]. Fig. 14 does so for the anisotropic Raman spectrum. We first consider the frequency region below 700 cm$^{-1}$. Although the LSC-IVR and classical MD demonstrate a qualitatively correct picture in the librational region (below 700 cm$^{-1}$) of the isotropic Raman spectrum (Fig. 13a), they produce poor results in the same region of the anisotropic Raman spectrum (Fig. 14a). The POLI2VS force field does not faithfully describe this region in the Raman spectroscopy of heavy water, similar to what we have discussed on the liquid water results.

We then focus on the Raman modes between 700 cm$^{-1}$ and the O-D stretching band. The LSC-IVR shows more satisfactory agreement with experiment than MD for the D-O-D bending band. The peak position of the D-O-D bending band produced by the LSC-IVR is blue-shifted by ~35 cm$^{-1}$ from experiment in the isotropic Raman spectrum and by ~50 cm$^{-1}$ in the anisotropic Raman spectrum, while that for MD is blue-shifted by ~53 cm$^{-1}$ in the isotropic spectrum and by ~73 cm$^{-1}$ in the anisotropic spectrum (as shown in Figs. 13a and 14a). Both the relative intensity and the structure of the spectrum are better reproduced by the LSC-IVR than by MD in the intermediate region between the librational and bending bands (700-1100 cm$^{-1}$) and also in the intermediate region between the bending and stretching bands (1300-2000 cm$^{-1}$). For instance, Fig. 13a shows that in the isotropic Raman spectrum the small peak around 786 cm$^{-1}$ and the three peaks around 1444 cm$^{-1}$, 1687 cm$^{-1}$, 1856 cm$^{-1}$ are semi-quantitatively captured by the LSC-IVR (although the peak positions and shapes are not perfectly reproduced) while these peaks are completely absent in the classical MD simulation. It implies that nuclear quantum dynamical effects are decisive in the Raman modes of these intermediate regions.

It is also expected that abundant quantum dynamical effects exist in the Raman modes of the O-D stretching band. The peak position of the O-D stretching band produced by the LSC-IVR is blue-shifted by ~3 cm$^{-1}$ from the experimental result in the isotropic spectrum (Fig. 13b) and by ~50 cm$^{-1}$ in the anisotropic spectrum (Fig. 14b), while that yielded by classical MD is blue-shifted by ~77 cm$^{-1}$ from experiment in the isotropic spectrum (Fig. 13b) and by ~111 cm$^{-1}$ in the anisotropic spectrum (Fig. 14b). In experiment the FWHM of the O-D stretching band is ~247 cm$^{-1}$ in the isotropic spectrum (Fig. 13b) and ~291 cm$^{-1}$ in the anisotropic spectrum (Fig. 14b). The LSC-IVR agrees reasonably well with

experiment—the LSC-IVR FWHM is ~214 cm$^{-1}$ in the isotropic spectrum (Fig. 13b) and ~260 cm$^{-1}$ in the anisotropic spectrum (Figs. 14b and 14c). In contrast, the FWHM produced by classical MD is ~174 cm$^{-1}$ in the isotropic spectrum (Fig. 13b) and ~243 cm$^{-1}$ in the anisotropic spectrum (Figs. 14b and 14c). While Fig. 13b shows that the peak at ~2384 cm$^{-1}$ of the experimental isotropic Raman spectrum is well reproduced by the LSC-IVR, the relative intensity of the shoulder in the higher frequency region of the stretching band (~2476 cm$^{-1}$) yielded by the LSC-IVR is noticeable but not as distinct as the experimental result. In the anisotropic Raman spectrum (Fig. 14b and Fig. 14c) the LSC-IVR yields a noticeable shoulder (~2384 cm$^{-1}$) in the lower frequency region of the O-D stretching band, although it is less distinguishable than the experimental result. In contrast, such shoulders are completely absent in the classical isotropic and anisotropic Raman spectra of heavy water.

## V.    Conclusion Remarks

The Raman spectroscopy offers a powerful tool to probe structure and dynamics of liquid water. Because the polarizability is a function of the instantaneous charge density and its response with respect to the electric field, the polarizability or polarizability-derivative is often a highly nonlinear function of coordinates or/and momenta. To study quantum dynamical effects in the Raman spectroscopy of liquid water it is then crucial to employ a good approximate quantum approach that is able to treat nonlinear operators well. Since the typical time scales related to the isotropic and anisotropic Raman spectra of liquid water are about 200-300 fs (relatively short), the inherent physical decay of this condensed phase system compensates the intrinsic unphysical decay of the LSC-IVR methodology. These suggest that the LSC-IVR can reasonably describe quantum dynamical effects in the isotropic and anisotropic Raman spectra of liquid water.

Our investigation verifies that abundant quantum dynamical effects exist in the stretching band of either of the isotropic and anisotropic Raman spectra. The LSC-IVR greatly improves over classical MD in reproducing both the peak position and lineshape of the experimental result. It is indeed encouraging that the LSC-IVR captures the two distinct peaks in the O-H stretching band of the isotropic Raman spectrum. In contrast, classical MD fails to present a qualitatively correct picture. This indicates that quantum dynamical effects play a critical role in describing the isotropic Raman modes in the stretching band.

Comparisons between the LSC-IVR and classical MD Raman spectra reveal that significant quantum dynamical effects also exist in the intermediate region between the librational and bending bands as well as in the intermediate region between the bending and stretching bands of the isotropic and anisotropic Raman spectra of liquid water (or heavy water) under ambient conditions. Quantum dynamical effects even play a decisive role in the two intermediate regions of the isotropic Raman spectrum—the peaks in these intermediate regions are well described in the LSC-IVR result but completely absent in the classical MD simulation of either liquid water or heavy water.

By selectively freezing either the intramolecular stretching mode or the intramolecular bending mode, we have first proposed two types of computer "experiments" to shed light on the Raman modes in the frequency region 1000-4000 cm$^{-1}$ of liquid water. The two types of computer "experiments" show that the coupling between the bending and stretching modes makes the bending band red-shifted and the stretching band blue-shifted. Such coupling is the source of the double-peak feature in the stretching band of the isotropic Raman spectrum, while it has very limited influence in the stretching band of the anisotropic Raman spectrum. It is confirmed that a substantial part of the peak in the intermediate region (2000-2400 cm$^{-1}$) of the anisotropic Raman spectrum should be attributed to the combined motion of the bending and librational modes. More interestingly, it is suggested that the peak in the intermediate region (2000-2400 cm$^{-1}$) of the isotropic Raman spectrum arises from the interplay of the stretching motion and the bending motion instead.

In the present paper we employ the POLI2VS water force field[115] to investigate the isotropic and anisotropic Raman spectra. Although the POLI2VS does not offer a satisfactory description for the low-frequency librational band as expected[115] (which may be consistently improved[112]), it does lead to reasonable and consistent quantum/semiclassical results for Raman modes in the higher frequency region (above ~1000 cm$^{-1}$ for liquid water and above ~700 cm$^{-1}$ for heavy water). The strategies that we use in the paper may be applied to study Raman spectra (or other types of vibrational spectra) of liquid water with other *ab initio*-based force fields[25, 95-106, 109, 110, 113] or *ab initio* calculations[107, 108, 111, 153]. E.g., it will be interesting in future to simulate the Raman spectroscopy of liquid water by using the LSC-IVR with the revPBE0-D3 DFT functional[107], MB-pol water force field[103, 105, 113, 114], or other promising PES/DMS/PTS for liquid water. It is expected that more accurate and consistent PES/DMS/PTS will lead to more reliable simulation results. It should also be stressed that an accurate representation of the underlying (Born-Oppenheimer) potential energy surface that reproduces structural and thermodynamic properties does not necessarily have a rigorous description of the linear IR and third-order Raman responses of the same system[40, 115, 154]. It is then crucial to employ quantum dynamics methods to test the accuracy of the PES/DMS/PTS for comparison to experiment.

Finally, we note that the numerical trick that we introduce in Eq. (33) and Eq. (34) is generally useful for the LSC-IVR to calculate other correlation functions involving scalar, vector, or tensor operators. As demonstrated for the Raman spectrum, we avoid calculating the derivative of the polarizability tensor over coordinates in Eq. (30) and in Eq. (31) by using the finite difference along the direction of a related vector. It greatly simplifies the part where the initial condition is involved in computing the LSC-IVR correlation function. It is also expected the strategy in Eq. (33) and Eq. (34) will be useful for such as path integral Liouville dynamics for evaluating the correlation function[155-157].

**Acknowledgement**

We thank Taisuke Hasegawa and Yoshitaka Tanimura for sending us the code for their POLI2VS water force field. This work was supported by the Ministry of Science and Technology of China (MOST) Grants No. 2016YFC0202803 and No. 2017YFA0204901, by the National Natural Science Foundation of China (NSFC) Grants No. 21373018 and No. 21573007, by the Recruitment Program of Global Experts, by Specialized Research Fund for the Doctoral Program of Higher Education No. 20130001110009, and by Special Program for Applied Research on Super Computation of the NSFC-Guangdong Joint Fund (the second phase) under Grant No. U1501501. We acknowledge the Beijing and Tianjin supercomputer centers and the High-performance Computing Platform of Peking University for providing computational resources. This research also used resources of the National Energy Research Scientific Computing Center, a DOE Office of Science User Facility supported by the Office of Science of the U.S. Department of Energy under Contract No. DE-AC02-05CH11231.

**Figure Captions**

**Fig. 1** (Color). The normalized local normal frequency distribution of liquid water at $T = 298.15$ K using the POLI2VS model.

**Fig. 2** (Color). Normalized polarizability-derivative correlation functions for liquid water at $T = 298.15$ K using the POLI2VS model. (a) The isotropic ones $\langle \hat{\dot{\alpha}}_{iso}(0)\hat{\dot{\alpha}}_{iso}(t) \rangle_{Kubo} / \langle \hat{\dot{\alpha}}_{iso}(0)\hat{\dot{\alpha}}_{iso}(0) \rangle_{Kubo}$. (b) The anisotropic ones $\langle \mathrm{Tr}[\hat{\dot{\boldsymbol{\beta}}}(0)\hat{\dot{\boldsymbol{\beta}}}(t)] \rangle_{Kubo} / \langle \mathrm{Tr}[\hat{\dot{\boldsymbol{\beta}}}(0)\hat{\dot{\boldsymbol{\beta}}}(0)] \rangle_{Kubo}$. (Statistical error bars included.)

**Fig. 3** (Color). The isotropic Raman spectrum of liquid water at 298.15 K using the POLI2VS model. The experimental results are adapted from Ref. [76]. (a) The low-frequency region. The relative intensity is magnified by a factor of 25. (b) The high-frequency region.

**Fig. 4** (Color). The anisotropic Raman spectrum of liquid water at 298.15 K using the POLI2VS model. The experimental results are adapted from Ref. [76]. (a) The low-frequency region. The relative intensity is magnified by a factor of 15. (b) The high-frequency region. (c) The line shape of the stretching band, where the LSC-IVR and classical MD results are shifted to the same peak position of the experimental result.

**Fig. 5** (Color). Same as Fig. 2, but for liquid water with constrained O-H bond lengths at $T = 298.15$ K using the POLI2VS model.

**Fig. 6** (Color). The isotropic Raman spectrum of liquid water with constrained O-H bond lengths at 298.15 K using the POLI2VS model. The experimental results are adapted from Ref. [76]. (a) Comparison of the LSC-IVR results to the experimental data in the low-frequency region. The relative intensity is magnified by a factor of 25. (b) Comparison of the LSC-IVR results to the experimental data in the high-frequency region. (c) Comparison of the classical results to the experimental data in the low-frequency region. The relative intensity is magnified by a factor of 25. (d) Comparison of the classical results to the experimental data in the high-frequency region.

**Fig. 7** (Color). Same as Fig. 6, but for the anisotropic Raman spectrum. The relative intensity in Panel (a) or (c) is magnified by a factor of 15.

**Fig. 8** (Color). Same as Fig. 2, but for liquid water with constrained H-O-H bond angles at $T = 298.15$ K using the POLI2VS model.

**Fig. 9** (Color). Same as Fig. 6, but for liquid water with constrained H-O-H bond angles at $T = 298.15$ K using the POLI2VS model.

**Fig. 10** (Color). Same as Fig. 7, but for liquid water with constrained H-O-H bond angles at $T = 298.15$ K using the POLI2VS model.

**Fig. 11** (Color). Same as Fig. 1, but for heavy water at $T = 298.15$ K using the POLI2VS model.

**Fig. 12** (Color). Same as Fig. 2, but for heavy water at $T = 298.15$ K using the POLI2VS model.

**Fig. 13** (Color). Same as Fig. 3, but for heavy water at $T = 298.15$ K using the POLI2VS model. The relative intensity in the low-frequency region [Panel (a)] is magnified by a factor of 20.

**Fig. 14** (Color). Same as Fig. 4, but for heavy water at $T = 298.15$ K using the POLI2VS model. The relative intensity in the low-frequency region [Panel (a)] is magnified by a factor of 10.

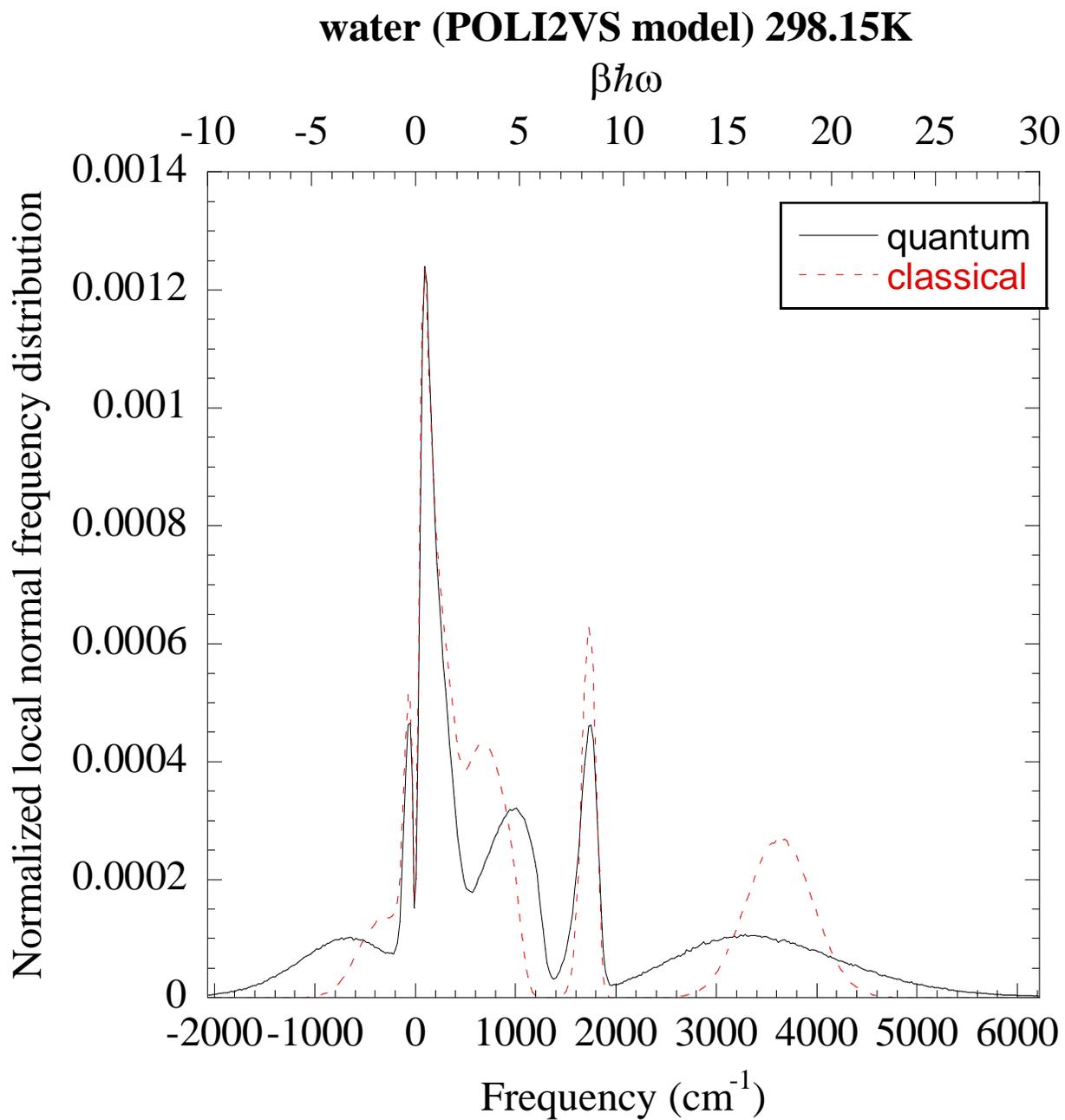

**Fig. 1**

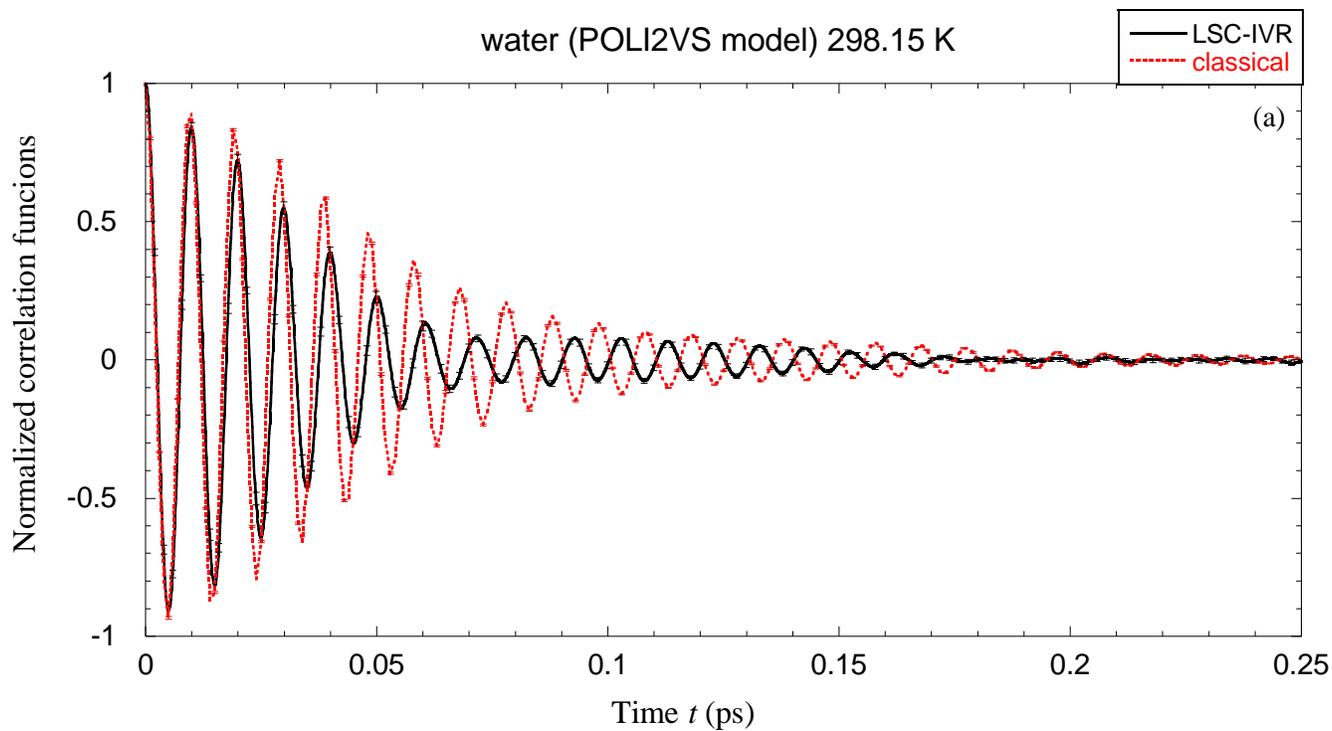

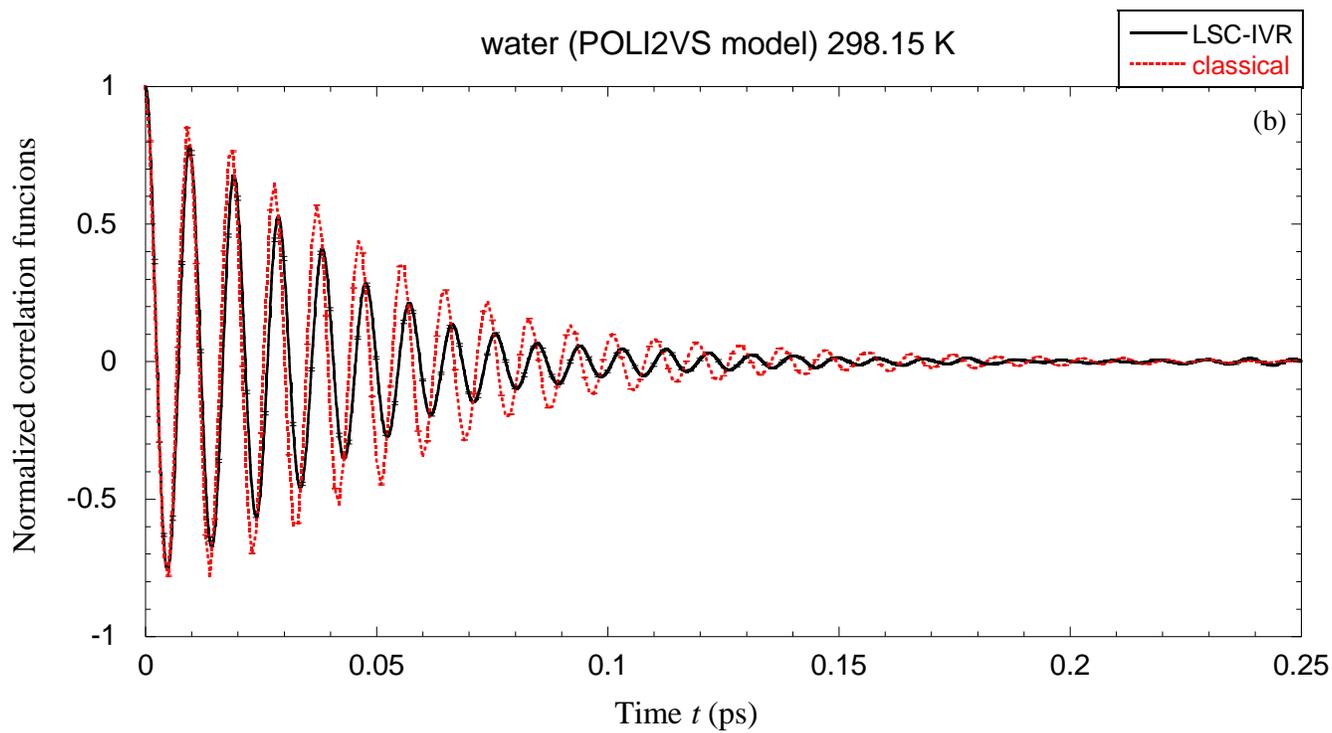

**Fig. 2**

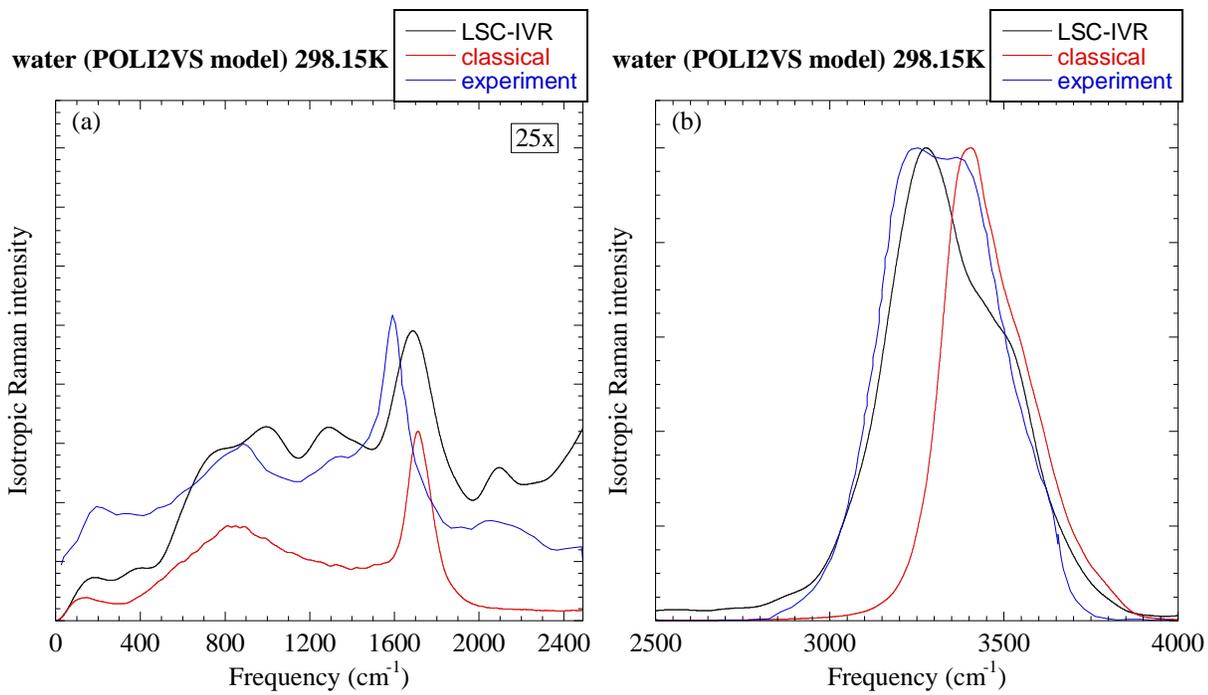

**Fig. 3**

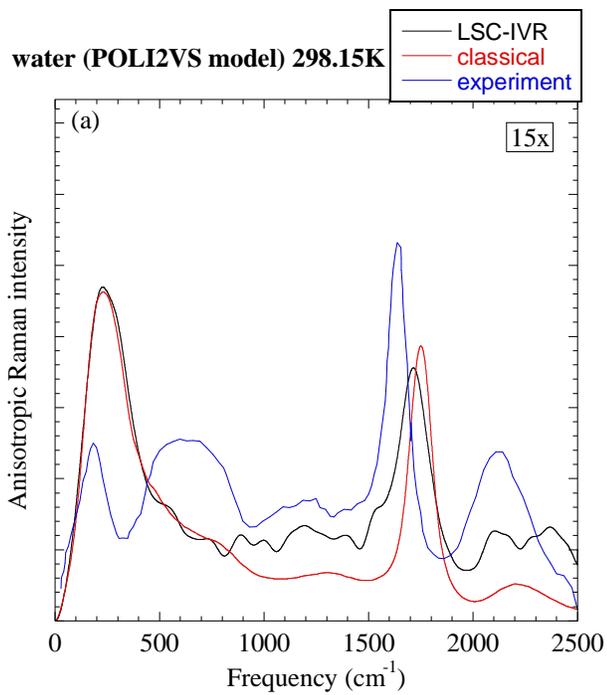
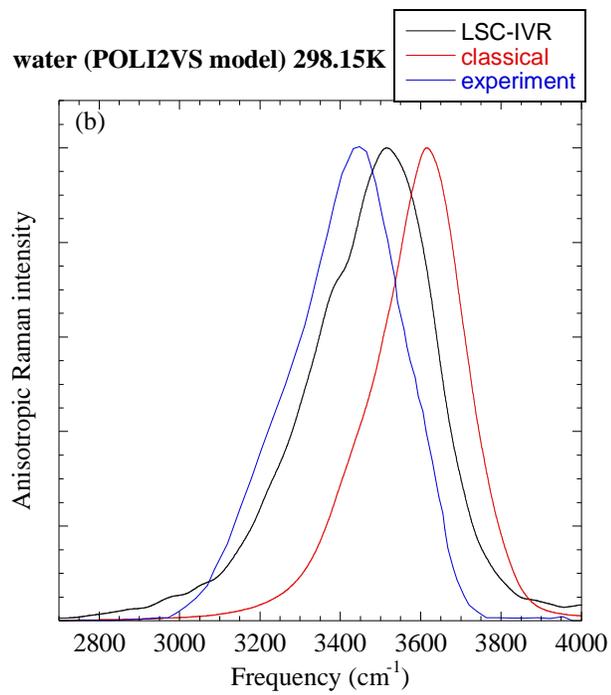
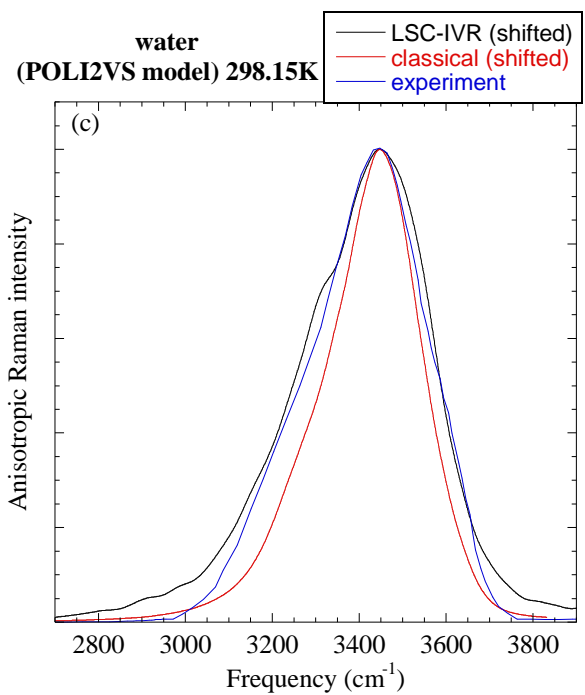

**Fig. 4**

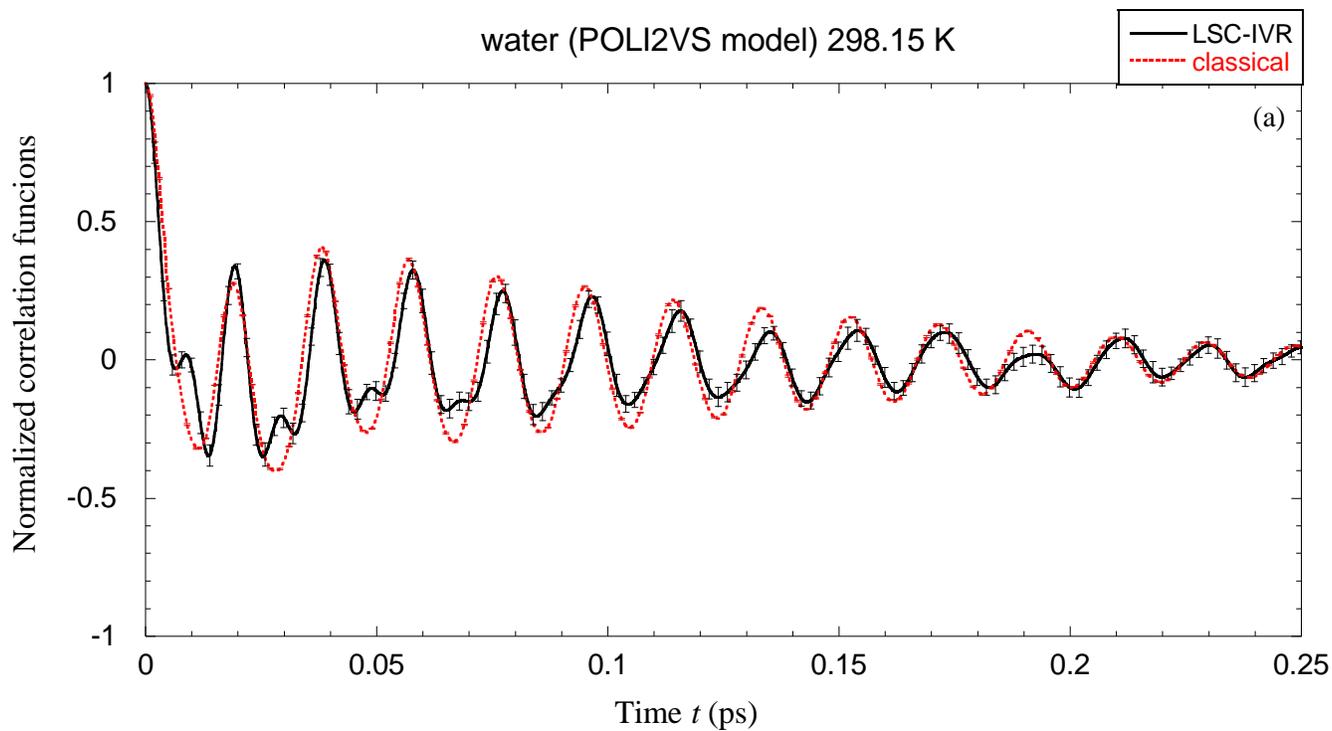

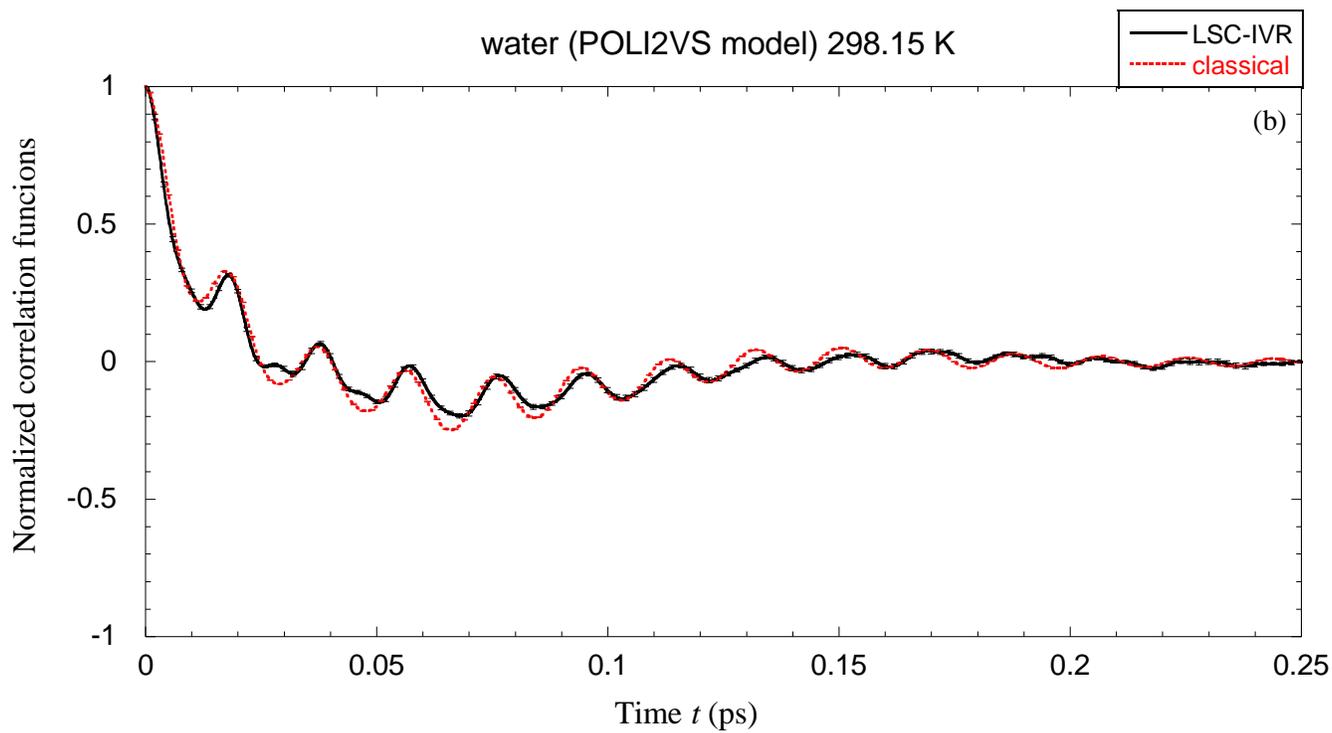

**Fig. 5**

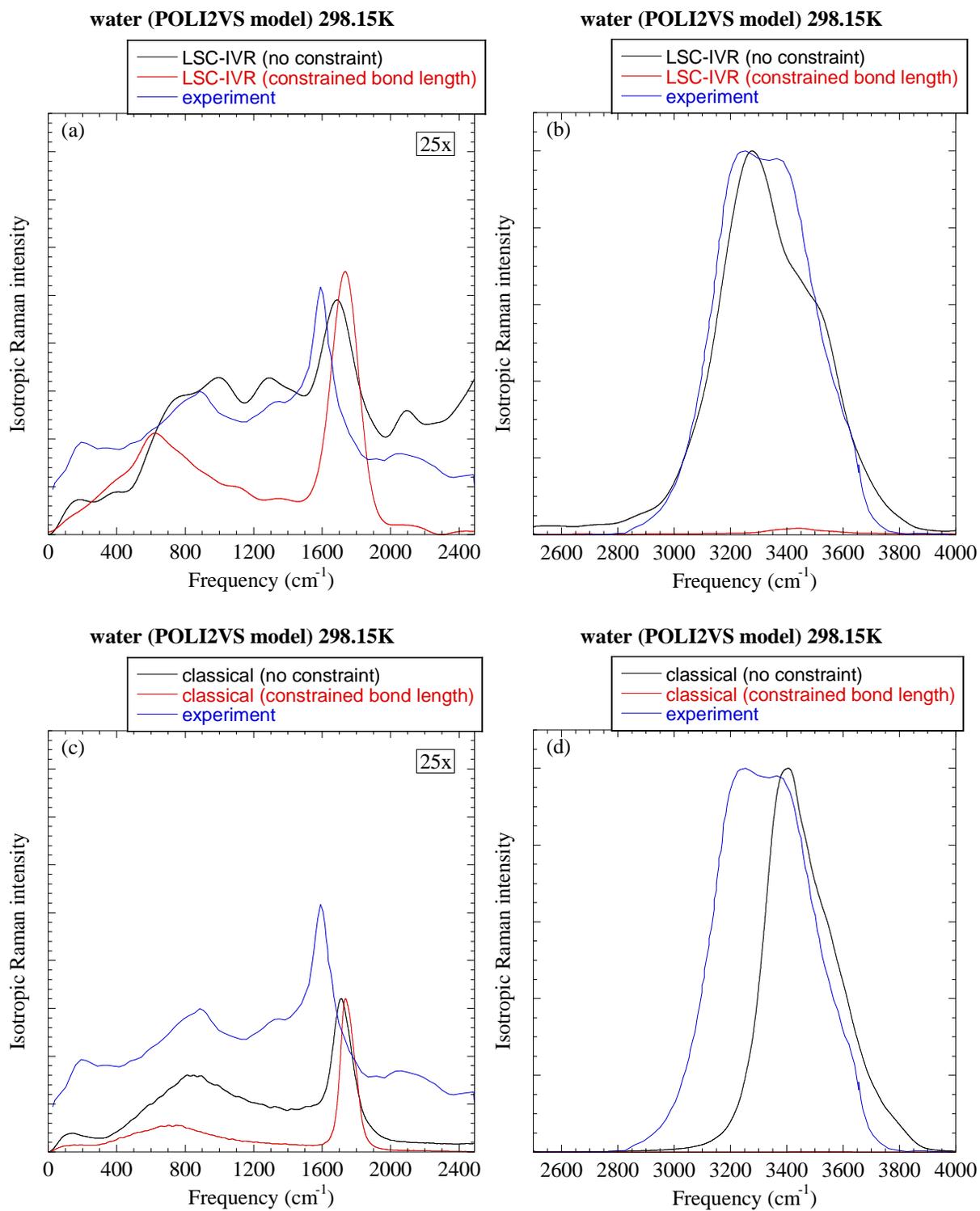

**Fig. 6**

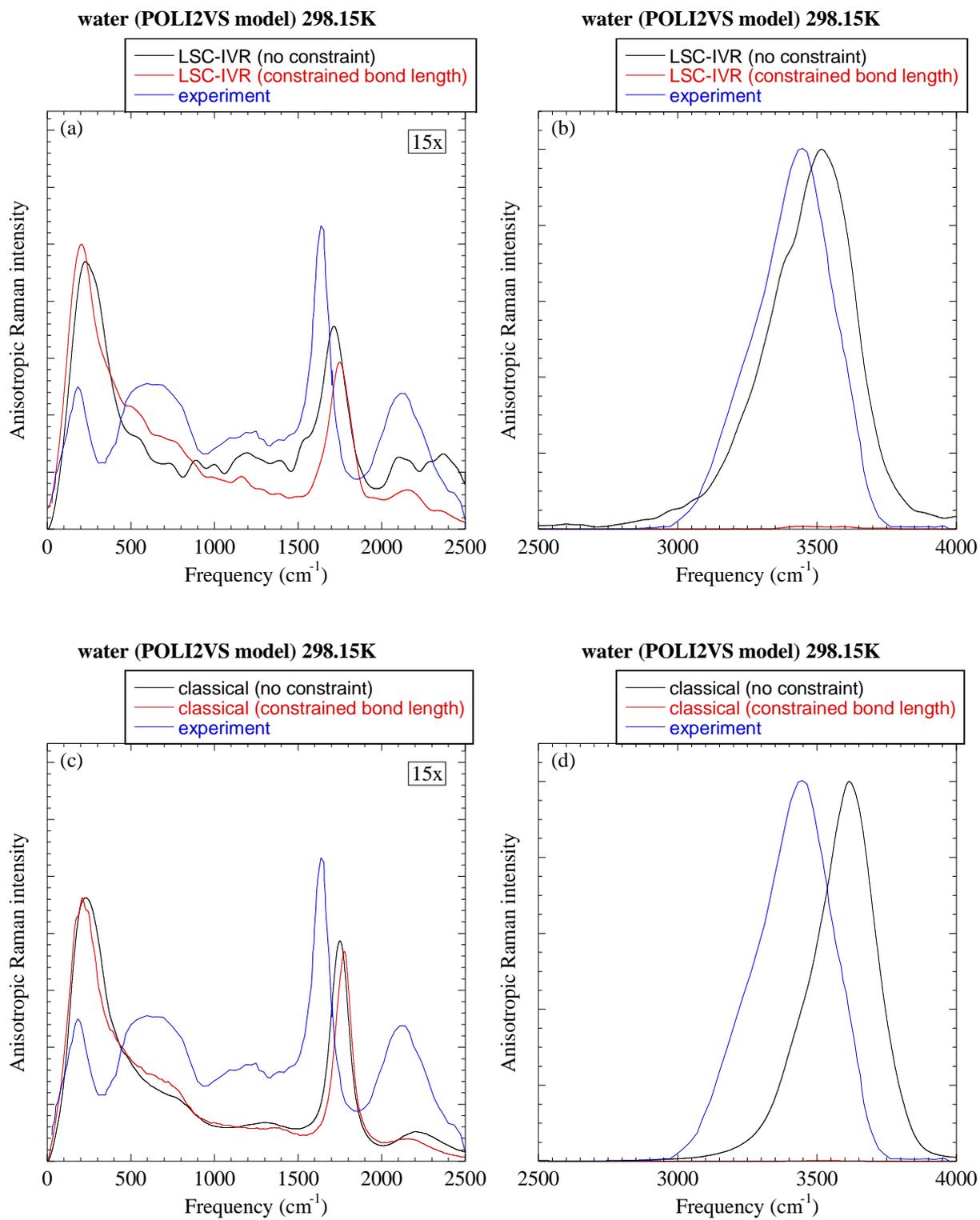

**Fig. 7**

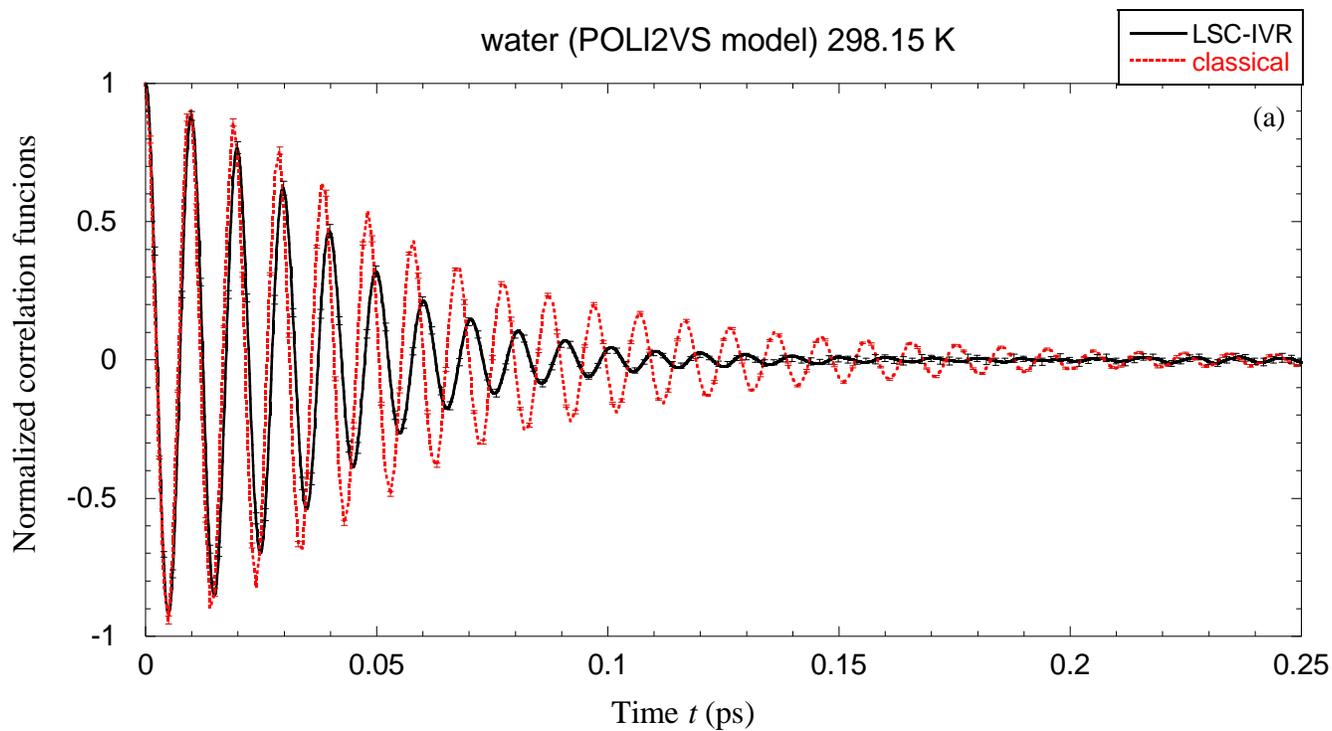

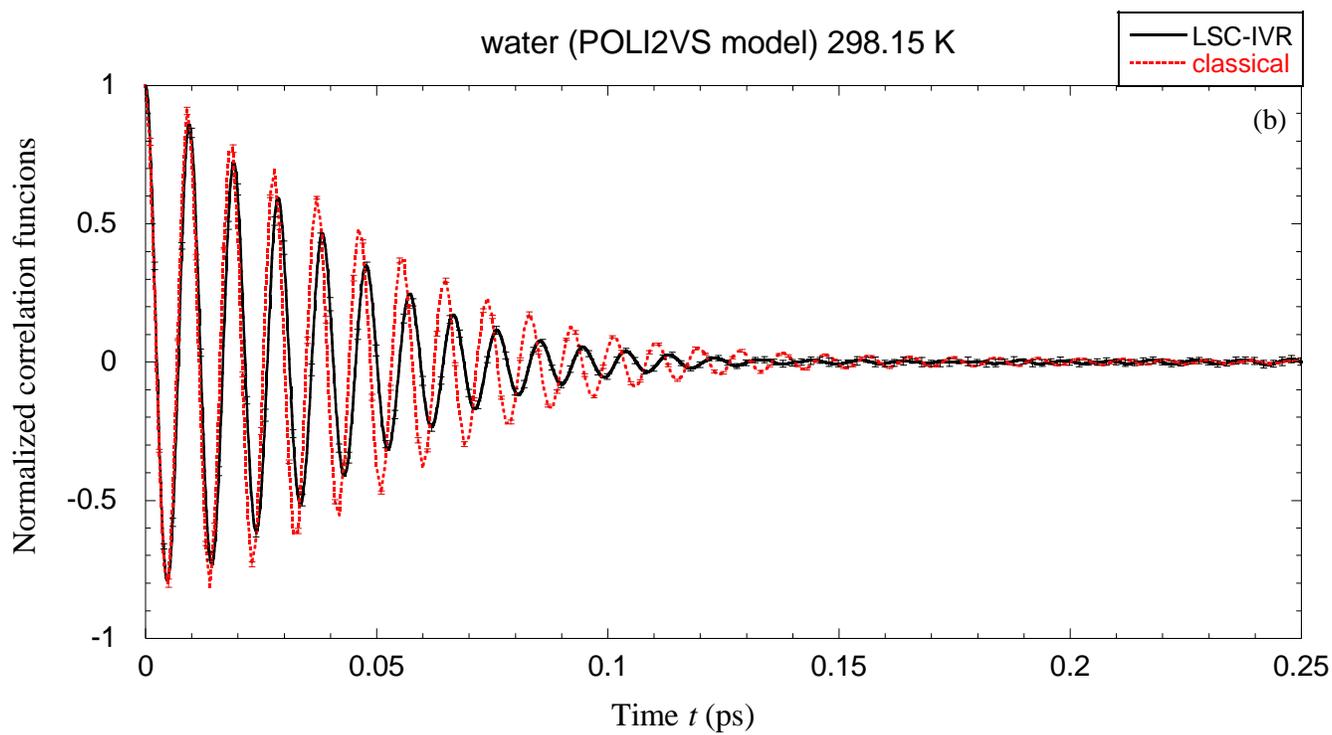

**Fig. 8**

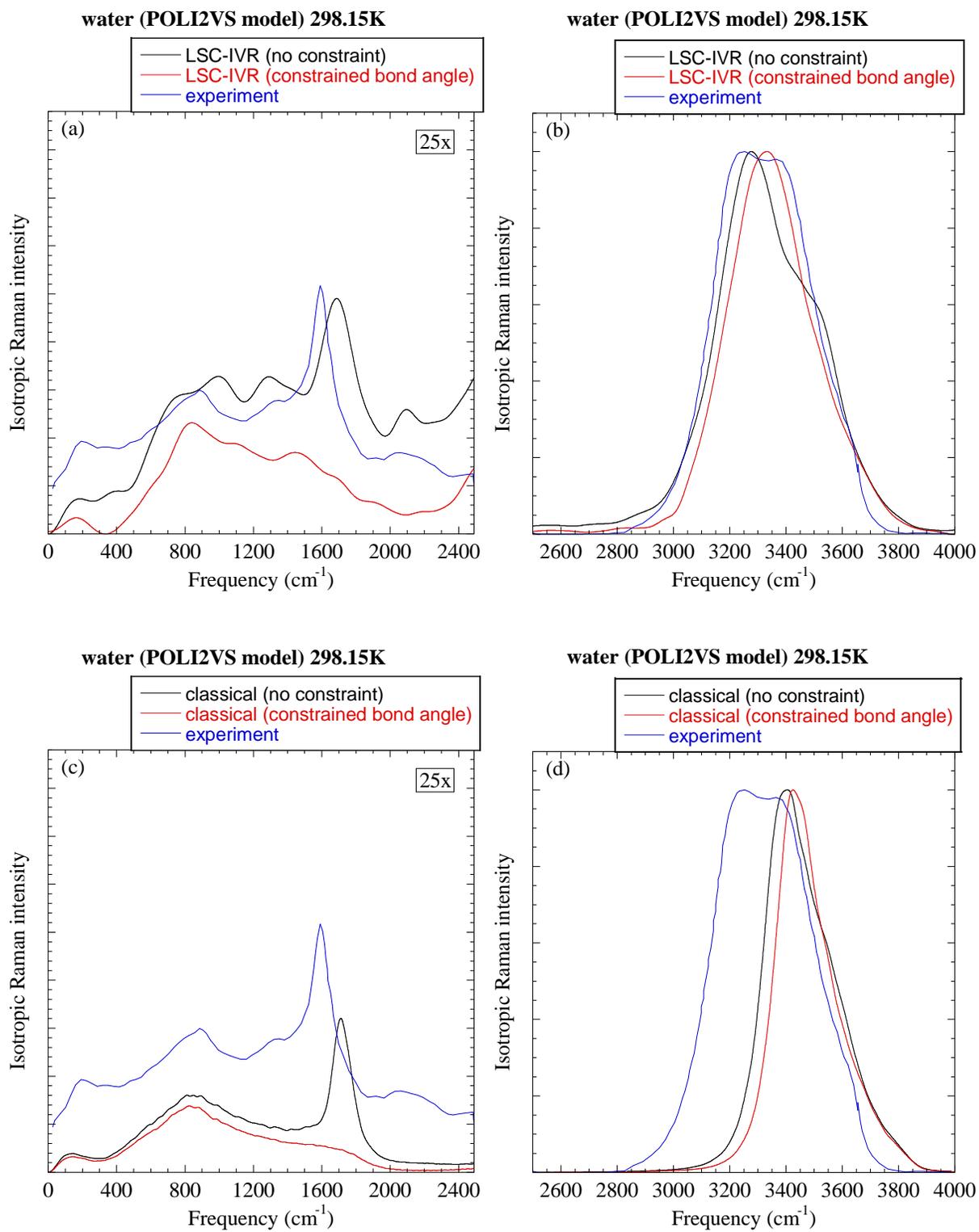

**Fig. 9**

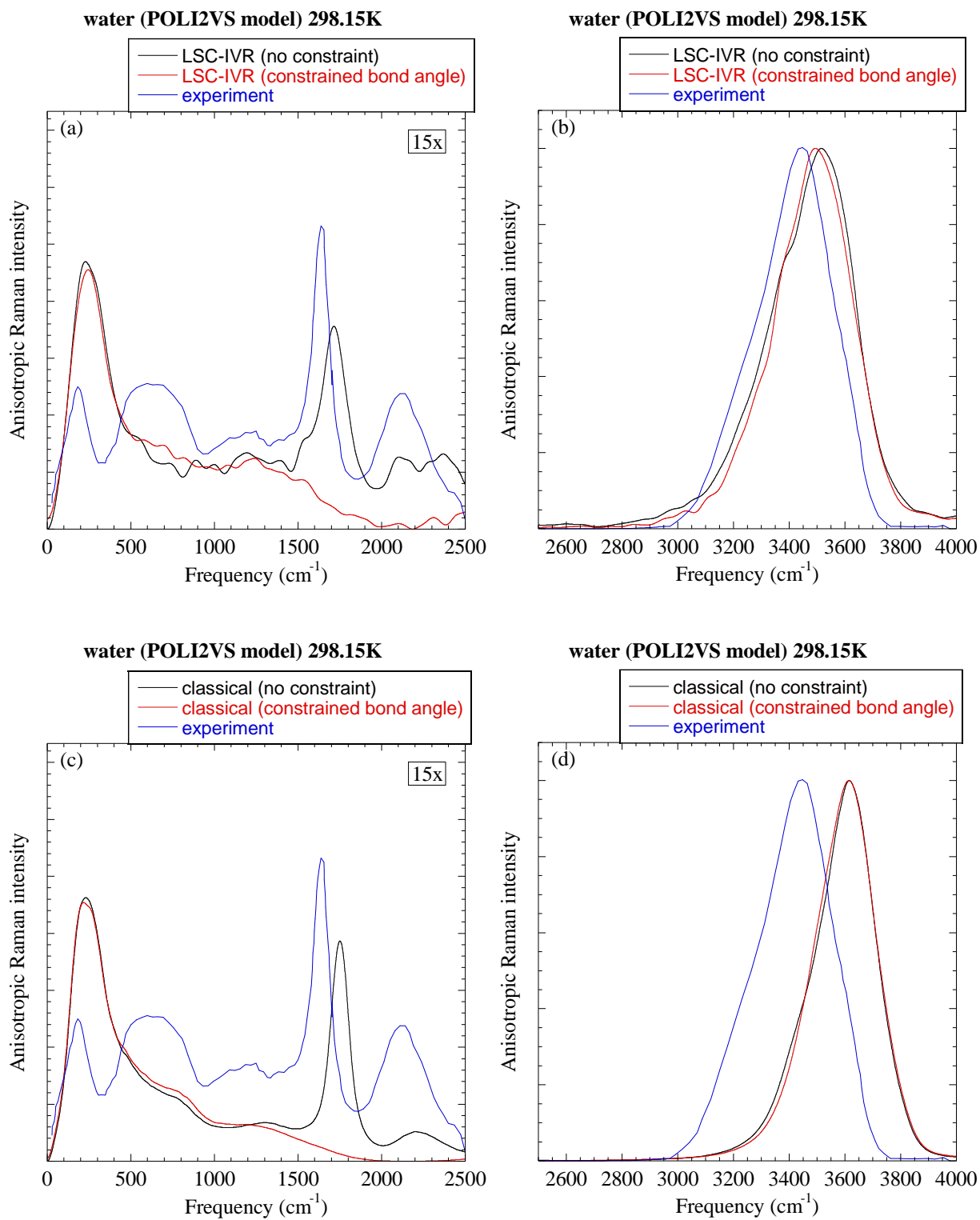

**Fig. 10**

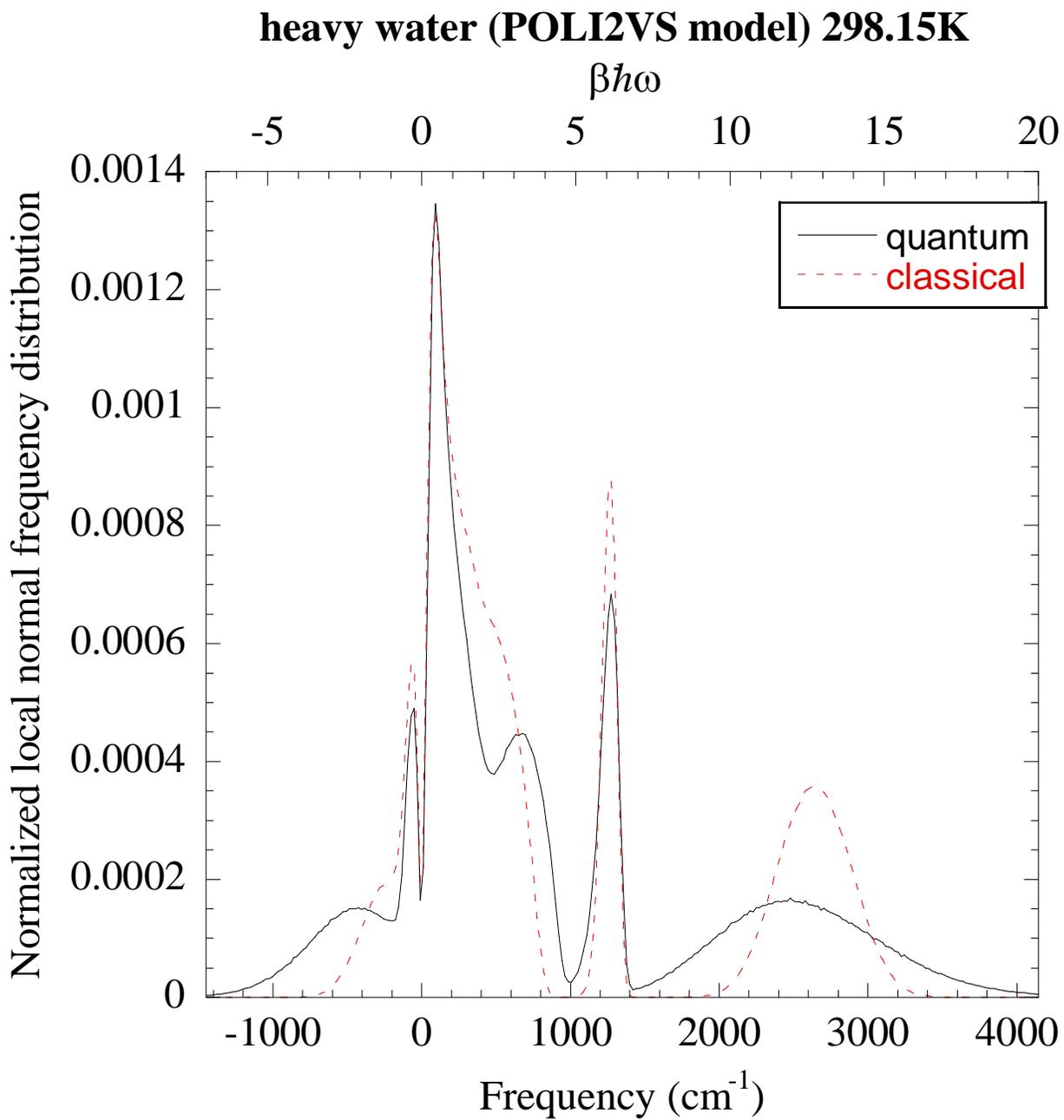

**Fig. 11**

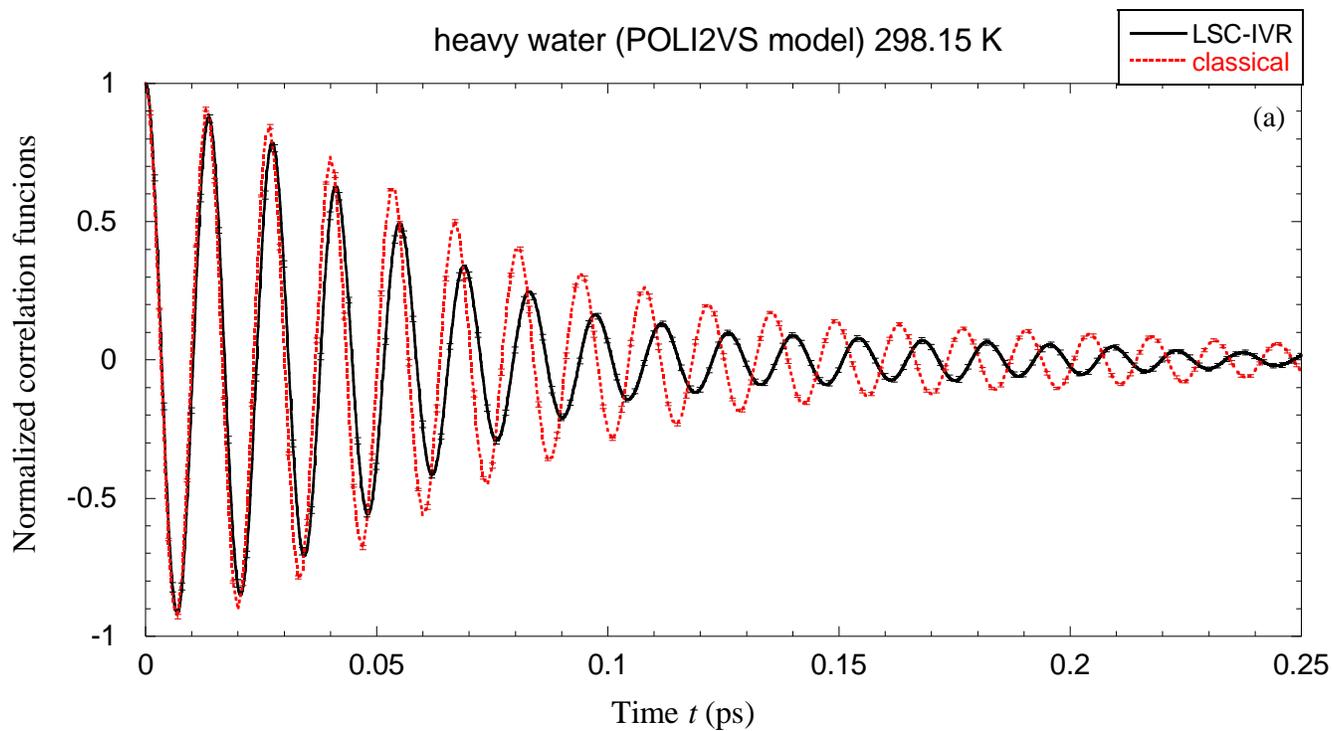
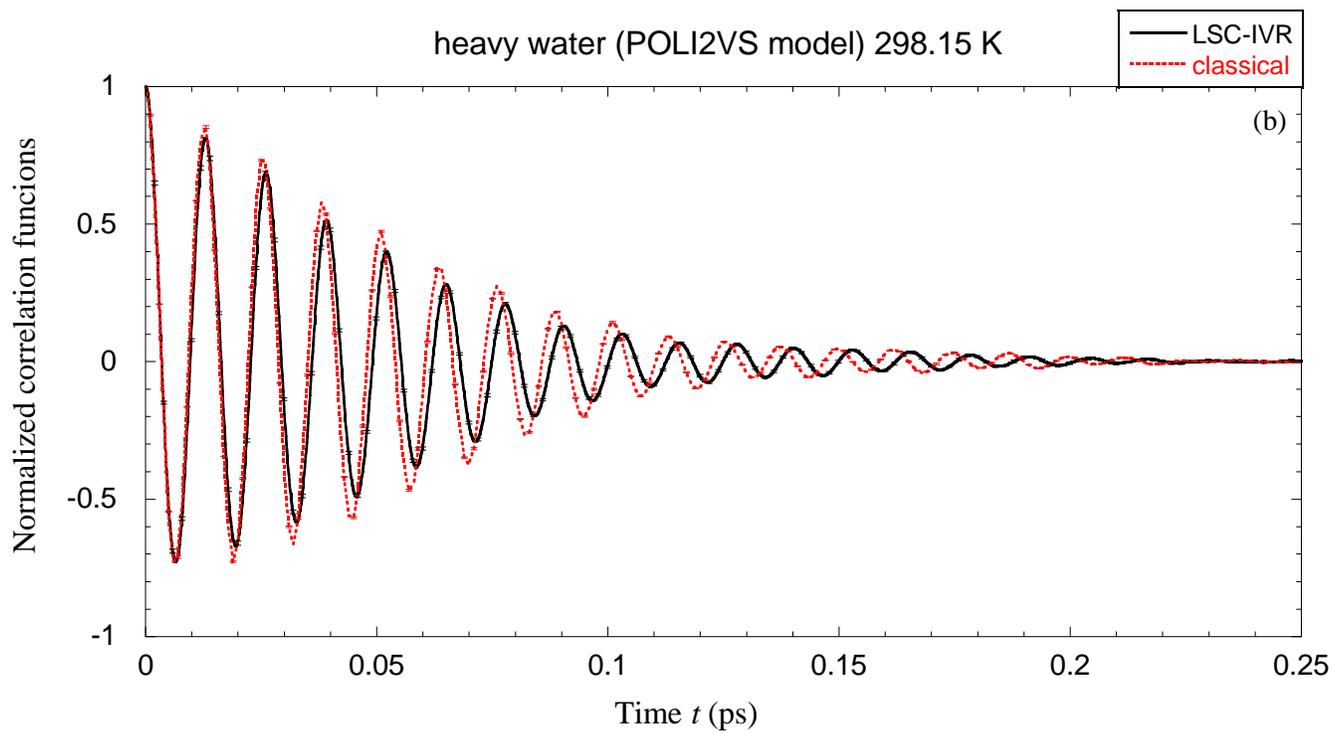

**Fig. 12**

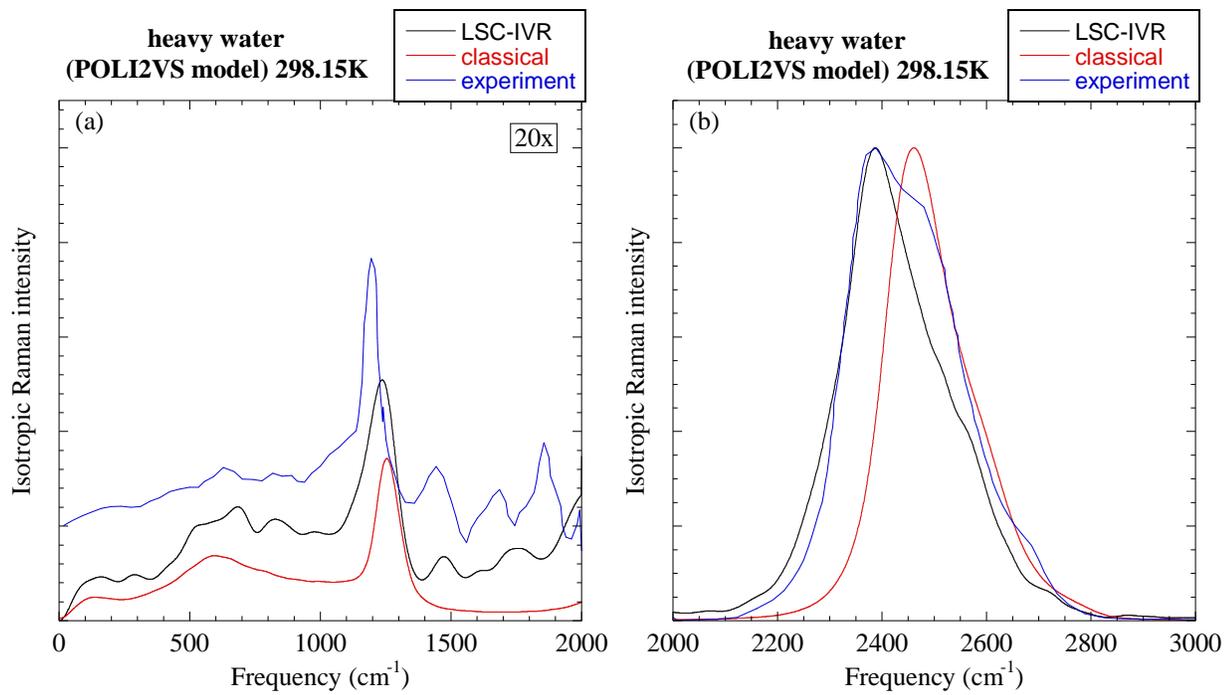

**Fig. 13**

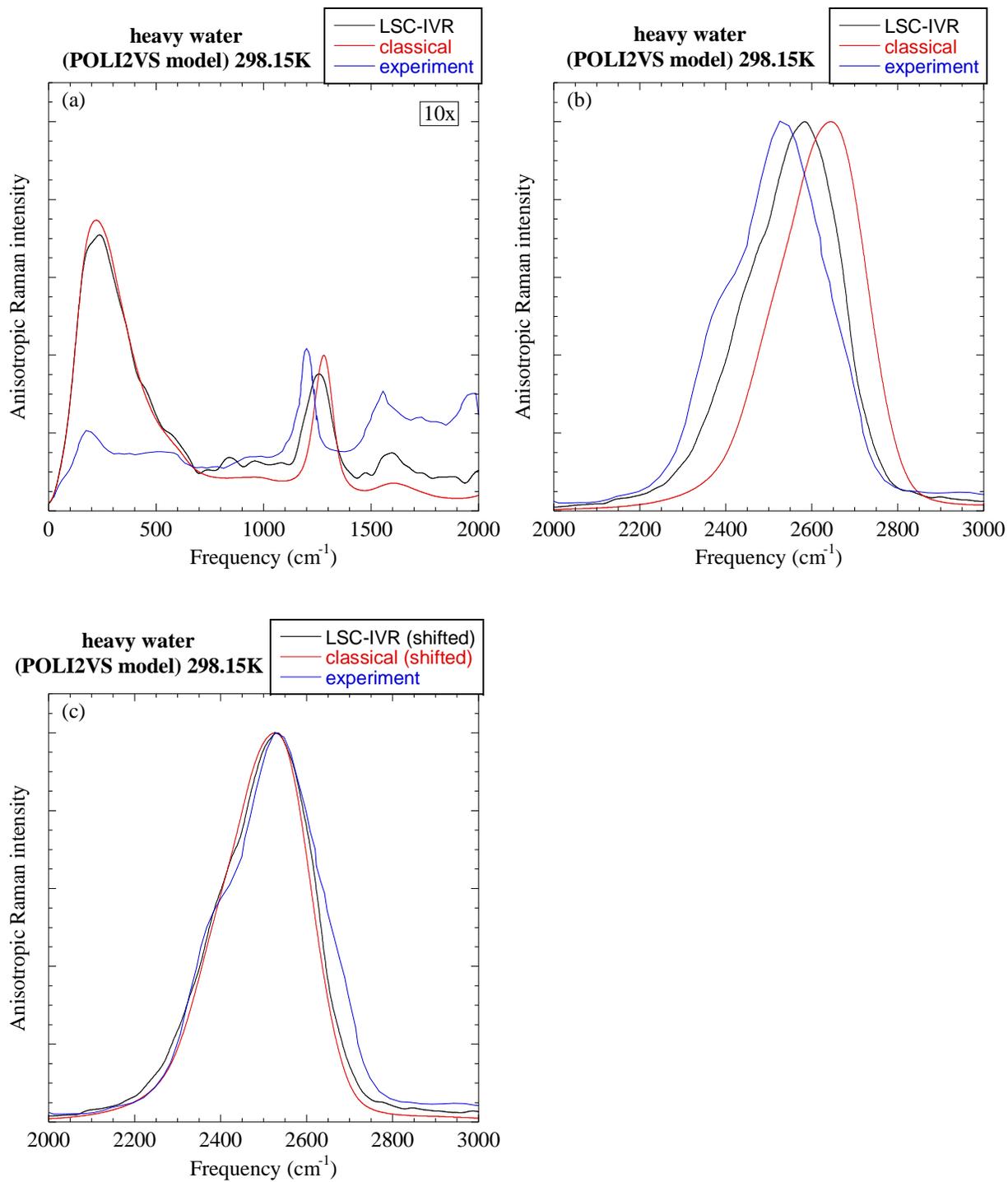

**Fig. 14**